\documentclass[preprint,12pt]{elsarticle}

\usepackage{xspace}
\usepackage{txfonts}
\usepackage{graphicx}

\usepackage{amssymb}

\journal{Astroparticle Physics}

\newcommand{\Fe}{\textit{Fermi}\xspace}
\newcommand{\He}{H.E.S.S.\xspace}
\newcommand{\Ve}{VERITAS\xspace}
\newcommand{\HeII}{H.E.S.S.~II\xspace}
\newcommand{\Ma}{MAGIC\xspace}
\newcommand{\MaII}{MAGIC-II\xspace}

\begin{document}

\begin{frontmatter}



\title{Potential of the next generation VHE instruments to probe the EBL (I): the low- and mid-VHE}


\author[MR]{Martin Raue\corref{cor1}}
\ead{martin.raue@desy.de}

\address[MR]{Institut f\"ur Experimentalphysik, Universit\"at Hamburg, Hamburg, Germany}
 
\author[DM]{Daniel Mazin}
\ead{mazin@ifae.es
}
\address[DM]{Institut de Fisica d'Altes Energies (IFAE), Edifici Cn. Universitat Autonoma de Barcelona, 08193 Bellaterra (Barcelona), Spain}

\cortext[cor1]{Corresponding author.}

\begin{abstract}
The diffuse meta-galactic radiation field at ultraviolet to infrared wavelengths - commonly labeled extragalactic background light (EBL) - contains the integrated emission history of the universe. Difficult to access via direct observations, indirect constraints on its density can be derived through observations of very-high energy (VHE; E$>$100\,GeV) $\gamma$-rays from distant sources: the VHE photons are attenuated via pair-production with the low energy photons from the EBL, leaving a distinct imprint in the VHE spectra measured on earth. Discoveries made with current generation VHE observatories like \He and \Ma enabled strong constraints on the density of the EBL, especially in the near-infrared. In this article the prospect of future VHE observatories to derive new constraints on the EBL density are discussed. To this end, results from current generation instruments will be extrapolated to the future experiment's sensitivity and investigated for their power to enable new methods and improved constraints on the EBL density.
\end{abstract}


\end{frontmatter}


\section{Introduction}\label{introduction}

The observation of very-high energy $\gamma$-rays (VHE; $E>100$\,GeV) from distant sources offers the unique possibility to probe the density of the meta-galactic radiation field at ultraviolet (UV) to infrared (IR) wavelengths, which is commonly labeled the extragalactic background light (EBL; typically 0.1-100\,$\mu$m). The VHE $\gamma$-rays interact with the low energy EBL photons via the pair production process ($\gamma_{\mathrm{VHE}} \gamma_{\mathrm{EBL}} \rightarrow e^+ e^-$) and the flux is attenuated \cite{nikishov:1962a,gould:1967a}. This attenuation can leave distinct signatures in the measured VHE spectra. With assumptions about the source physics and the spectrum emitted at the source location (intrinsic spectrum), constraints on the density of the EBL can be derived \cite[e.g.][]{stecker:1994a, dwek:1994a}.

The current generation of VHE instruments (\He, \Ma, \Ve) significantly increased the number of known extragalactic VHE sources from 4 in the year 2003 to more than 25 today.  These discoveries, combined with the advanced spectral resolution of these instruments and the wide energy range they cover, led to new strong constraints on the EBL density, in particular at optical to near-IR (NIR)  wavelengths \cite{aharonian:2006:hess:ebl:nature,mazin:2007a,albert:2008:magic:3c279:science}. Since these limits depend on assumptions about the source physics, the strong constraints also sparked intense discussions on the validity of the assumptions and possible caveats \cite[e.g.][]{aharonian:2002c, katarzynski:2006a, reimer:2007a,aharonian:2008a,krennrich:2008a,raue:2009b}. This discussion has not yet converged and there are interesting arguments for both sides.

Current generation systems have recently been upgraded (\MaII) or the upgrades are under construction (\HeII). These upgrades are mainly aimed to improve the overall sensitivity by a factor two to three and extend the energy range toward the lower energy regime of 20 to 100\,GeV. This will lead to some improvements, but a quantitative difference or a breakthrough compared to the performance of the existing facilities will only be achieved with an order of magnitude improvement in sensitivity. The Next Generation Cherenkov Telescope Systems (NGCTS) are in the advanced planing phase aiming to achieve this order of magnitude improvement: the Cherenkov Telescope Array (CTA\footnote{http://www.cta-observatory.org/}) \cite{hermann:2007a} and the Advanced Gamma-ray Imaging System (AGIS\footnote{http://www.agis-observatory.org/}) \cite{buckley:2008a}. Whereas CTA envisions to improve the sensitivity over a wide energy range from the few tens of GeV to the multi TeV regime, AGIS mainly concentrates on energies above 100 GeV, an extended field of view and an improvement of the angular resolution.

The potential of these upcoming experiments to probe the EBL is the topic of this article. While an order of magnitude improvement in sensitivity for an astronomical instrument will always lead to new and unexpected results, this article will - as a first step - focus on known results and their extrapolation according to the sensitivities of the next generation instruments. Emphasis will be on new techniques enabled by the performance features (extended sensitivity and energy range) of the upcoming instruments.

For the calculations in the paper a standard $\Lambda$CDM cosmology with $h=\Omega_\Lambda=0.7$ and $\Omega_M=0.3$ is adopted.

\section{Basic assumptions and simulation details}\label{simdetails}
 
\begin{figure}[tbp]
\centering
\includegraphics[width=0.9\textwidth]{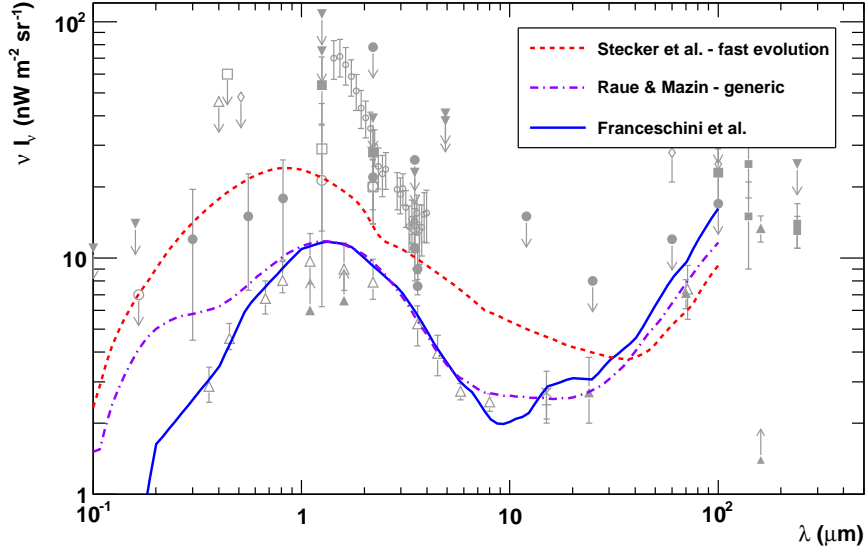}
\caption{Present day ($z = 0$) EBL density of the EBL models/shapes utilized in this study (Stecker et. al: \cite{stecker:2006a}; Raue \& Mazin:  \cite{raue:2008b}; Franceschini et al. \cite{franceschini:2008a}). Grey markers show measurements and limits on the EBL density (from \cite{mazin:2007a}).}
\label{Fig:EBLModels}
\end{figure}

\begin{figure}[tbp]
\centering
\includegraphics[width=0.75\textwidth]{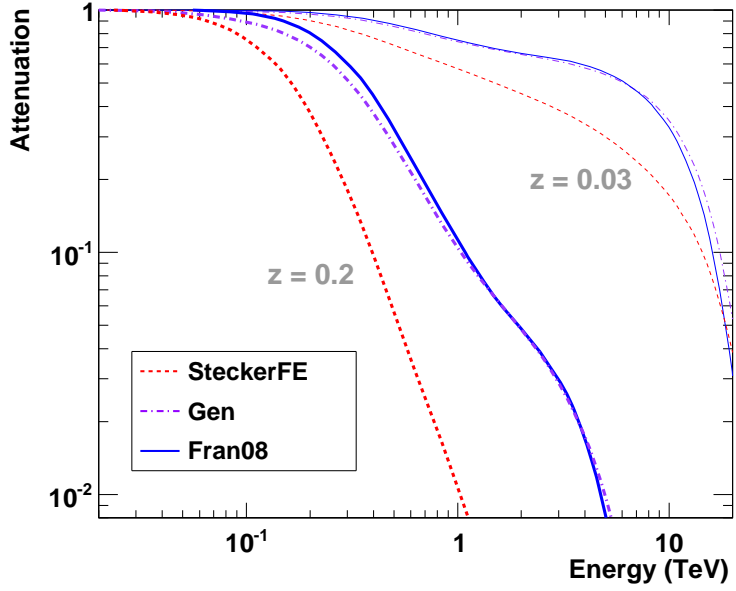}
\caption{Attenuation for VHE $\gamma$-rays for the EBL models utilized in this study for sources at redshift $z=0.03$ (thin lines) and $z=0.2$ (thick lines).}
\label{Fig:EBLAttenuation}
\end{figure}

\paragraph{EBL models and attenuation}
The precise level of the EBL density is not well known. Solid lower limits from integrated deep galaxies counts at optical and infrared wavelengths do exist \cite[e.g.][]{madau:2000a,fazio:2004a,frayer:2006a,dole:2006a}, but direct measurements are hampered by dominant foregrounds \cite{hauser:1998a}. Over a large wavelength region in the infrared the best upper limits on the EBL density are derived from VHE observations of distant sources \cite[e.g.][]{aharonian:2006:hess:ebl:nature,mazin:2007a,albert:2008:magic:3c279:science}. To account for the uncertainty in the EBL density two different approaches are followed here: (i) To illustrate the effect of the different EBL densities two extreme EBL models are used: for the low EBL density case the model from \cite{franceschini:2008a} (Fran08 in the following) is adopted, for the high model the fast evolution case from \cite{stecker:2006a} (SteckerFE in the following) is used. It should be noted that the later model is already disfavored by several VHE observations. (ii) Detailed studies of the effect of different EBL densities are carried out by scaling the EBL density presented in \cite{raue:2008b} (Gen in the following) and Fran08. The present day EBL densities ($z=0$) for the EBL models utilized in this work are displayed in Fig.~\ref{Fig:EBLModels}

Fig.~\ref{Fig:EBLAttenuation} displays the resulting attenuation for VHE $\gamma$-ray sources at 2 different redshifts ($z=0.03$ and $0.2$) for the models utilized. Several features can be identified:
\begin{itemize}
\item At low energies ($<80$\,GeV) the spectrum is practically not attenuated. Since this energy range will be sampled by the NGCTSs with high precision, it will be possible to measure the unabsorbed spectrum.
\item At energies between 80\,GeV and 2\,TeV the attenuation is increasing due to the EBL photons in the optical to near-infrared range peak of the EBL density.
\item At energies between 2 to 10\,TeV a flattening of the attenuation is expected, due to the $\sim \lambda^{-1}$ behavior of the EBL density in the near to mid-infrared, resulting in a constant attenuation. Such a modulation of the EBL attenuation has been considered as a possible key signature for EBL attenuation \cite[e.g.][]{aharonian:2002a,costamante:2003a}. Unfortunately, the intrinsic weakness of the sources combined with the sensitivity of the instruments make it very difficult to probe such a feature with previous or current generation experiments. For the SteckerFE model the EBL density in this wavelength range is flatter, resulting in a smoother attenuation from 100\,GeV to 10\,TeV, suppressing such a feature.
\item At energies around 10\,TeV the turnover in the EBL density towards the far-infrared peak of the EBL results in a strong attenuation, effectively resulting in a cut-off in the measured spectra.
\end{itemize}
The strength and the position of these features vary with the distance of the VHE sources, the assumed EBL model, and the overall EBL density.

\paragraph{EBL limits from VHE observations}
So far, VHE sources used to derive limits on the EBL density belong to a single source class, active galactic nuclei (AGNs), and the majority of them to the Blazar sub-class, which are AGNs with strong jet activity and the jets are closely aligned to the line of sight of the observer. Up to now, mainly two different methods - and thereby assumptions about the source intrinsic spectrum - have been utilized to derive limits on the EBL density:
\begin{itemize}
\item \textit{Spectral concavity.} It is assumed that the overall intrinsic source spectrum at high energies will follow a concave shape, or at least will not show an exponential rise towards the highest energies. These assumptions are well motivated by the common leptonic modeling of the sources under investigation (blazars), although different (maybe more exotic models) can possibly reproduce such a feature \cite[e.g.][]{aharonian:2002c}. Limits on the EBL density are derived by excluding EBL densities that would lead to such features in the observed sources. This method naturally probes the EBL at wavelengths from the mid to the far-infrared.
\item \textit{Maximum spectral hardness.} To probe the EBL in the optical to near-infrared, it is assumed that the intrinsic source spectrum cannot exceed a certain absolute hardness. While somewhat similar in spirit to the first method the underlying assumptions are stronger, since in the energy range of interest (100\,GeV to several TeV) the spectral shape of the intrinsic spectrum is more uncertain. While most of the basic models used to describe the source spectra indeed imply that the VHE spectrum does not exceed a certain hardness, the absolute value is less certain\footnote{e.g. $\Gamma = 1.5 - 0.6$ for $dN/dE \sim E^{-\Gamma}$ and simple leptonic models.} and possible source intrinsic effect (e.g. internal absorption \cite{aharonian:2008a}) could complicated the situation.
\end{itemize}

In this paper two different methods to derive limits on the EBL density will be explored: (i) utilizing the unabsorbed part of the VHE spectrum and (ii) searching for attenuation modulation signatures. While not completely new, it will be shown that with the NGCTS's extended energy range paired with its vastly improved sensitivity it will be possible to utilize these methods effectively for the first time. Method (i) holds the potential to derive limits on the EBL density with a minimal set of assumptions, while method (ii) enables to not only derive upper limits on the EBL density but to probe the absolute level.

\begin{figure}[tbp]
\centering
\includegraphics[width=0.8\textwidth]{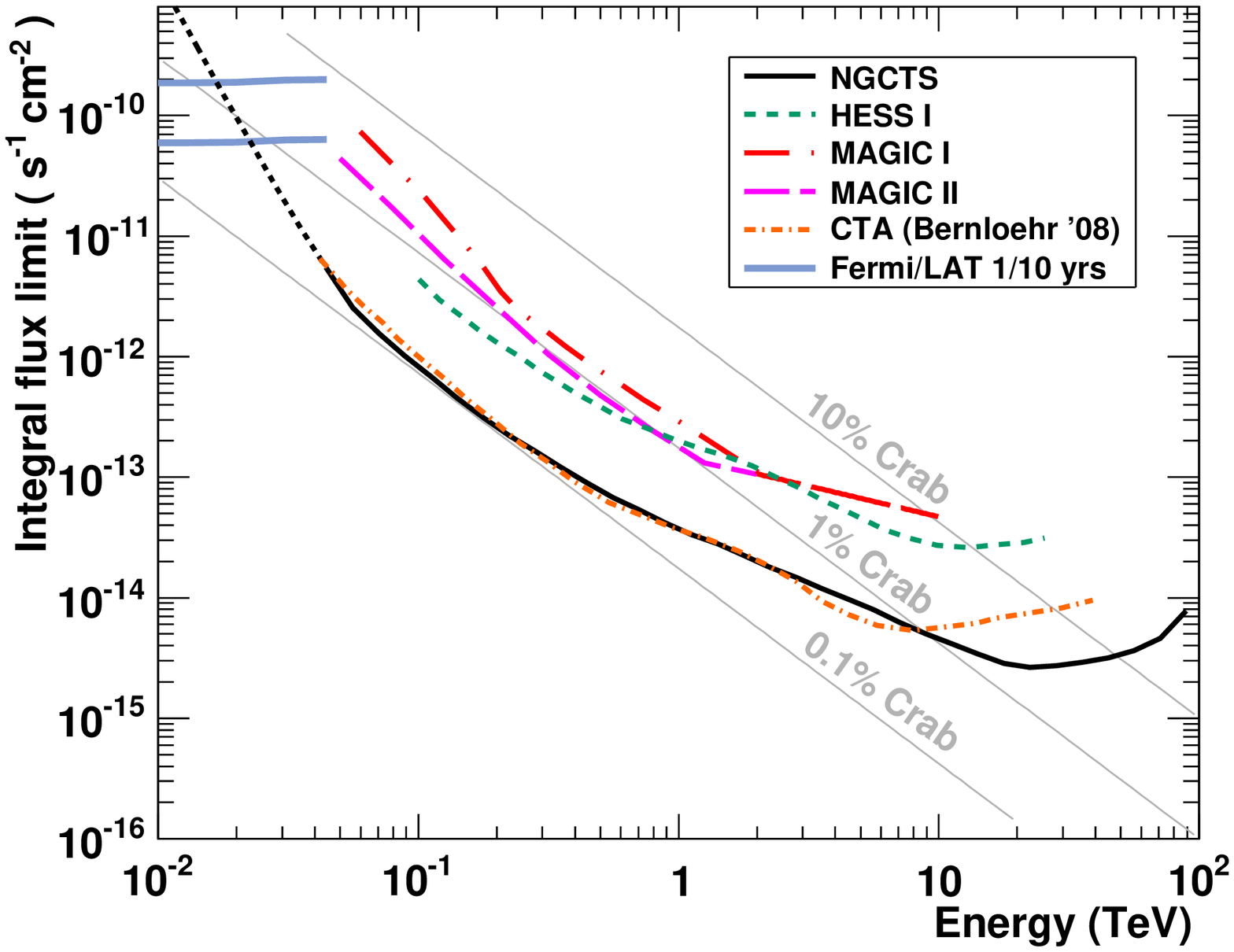}
\caption{Integral flux sensitivity ($5\sigma$ in 50\,h) of the NGCTS used in this study in comparison to the sensitivity of existing observatories (\He: \cite{bernloehr:2008a}; \Ma: \cite{schweizer:2009a}; \Fe: \texttt{http://www-glast.slac.stanford.edu/software/IS/glast\_lat\_Performance.htm}).}
\label{Fig:Sensitivity}
\end{figure}
 
\paragraph{Effective area, background rate and sensitivity}
To simulate the spectra, which will be detected by the NGCTS, assumptions about the sensitivity of the instrument have to be made. As baseline sensitivity, the sensitivity of the  "4 large + 85"  CTA array presented in \cite{bernloehr:2008a} is taken and the sensitivity is derived assuming an effective detector area (effective area) and a background rate. For the effective area the post cut MAGIC effective area at 20\,deg zenith angle (\cite{albert:2008:magic:crab}) scaled up by a factor 20 is adopted to reach an effective area exceeding $10^{6}$m$^2$ at energies above a few hundred GeV. In addition, the MAGIC effective area is shifted by a factor of 2 to lower energies to reflect the improved sensitivity at low energies.
\footnote{The energy threshold approximately scales linearly with 1/Area of the mirror area of the single telescope. For a CTA type NGCTS instruments the largest telescopes are expected to have mirror diameter in the order of 24 m which gives a factor two larger area compared to 17m for the MAGIC mirror.}
The resulting effective area $A_{NGCTS}$ versus energy $E$ in the energy range 20\,GeV to 20\,TeV is well described by the function
\begin{equation}
A_{NGCTS} = 2.5 \cdot 10^{16} \mbox{m}^2 (E/1\,\mbox{TeV})^{6} (1 + (E/(1\,\mbox{TeV} \cdot 0.02))^{0.9})^{-6/0.9}
\end{equation}
The differential background rate after event selection cuts is approximated by a broken power law function with photon index 3.6 below and 2.7 above the break energy of  300\,GeV, similar to what is seen in current generation instruments. The absolute level of the background flux rate is chosen so that the resulting sensitivity matches the sensitivity of the CTA "4 large + 85" array from \cite{bernloehr:2008a}.
The integral sensitivity adopted in this study for the NGCTS in comparison to current generation and future HE/VHE instruments is displayed in Fig.~\ref{Fig:Sensitivity}. In the overlap region it follows very well the CTA sensitivity for the "4 large + 85"  array from \cite{bernloehr:2008a} up to 10\,TeV. Compared to current instruments (\He, \Ma, \Ve) the sensitivity in the core energy range between 100\,GeV and 10\,TeV is improved by about one order of magnitude. In addition, the energy range is extended toward lower and higher energies. With the chosen parameters for the effective area and the background rate at low energies significant sensitivity is reached down to energies below 20\,GeV, which enables a large overlap region in energy with the Large Area Telescope (LAT) onboard \Fe satellite. For this energy region, where there is no overlap with the published CTA sensitivity ($<$40\,GeV), there is of course a certain degree of freedom in the choice of parameters and therefore the sensitivity is not very well constrained. In this study only energies between 40\,GeV and 10\,TeV will be used.

\paragraph{Spectrum simulation method}
To calculate the simulated spectrum the number of $\gamma$-photon events $N_{S}$ (signal) and background events $N_{BG}$ per energy bin need to be determined: the NGCTS effective area folded with the assumed intrinsic source flux is integrated over the energy bin; the same is done for the background events integrating over the background rate in the bin. Both numbers are multiplied with the effective observation time. Different functions are utilized to describe the intrinsic flux and they will be discussed further in the section where they first appear.
The attenuation of the source flux due to the EBL is calculated following the recipe given in \cite{mazin:2007a,raue:2009b}: the attenuation is directly folded into the intrinsic spectrum and then the attenuated intrinsic flux function folded with the effective area is integrated over the energy bin.
The number of signal events is randomly varied assuming a Poisson distribution. It is assumed that the background is well determined (e.g. via background measurements in a large sky area compared to the signal region).
The error on the number of photons $N_\sigma$ in a bin is derived utilizing equation 17 from \cite{li:1983a} assuming 5 background regions (i.e. an alpha factor of 0.2). The signal in an energy bin is considered significant when all of the following criteria are met: (1) the signal significance exceeds 3 standard deviations, (2) there are at least 10 excess events in the bin, (3) the number of excess events in the bin exceeds 3\% of the number of background events. This last condition takes into account a systematic error in the determination of the number of background events. These are rather conservative assumptions, since e.g. in current publications on VHE $\gamma$-astronomy often energy bins with significances down to 1.5\,$\sigma$ or less are included in the analysis.

IACT  experiments have a limited energy resolution which, for current  generation instruments, is in the order of $<$15\% for energies above 100\,GeV \citep[e.g.][]{aharonian:2006:hess:crab} and $<$40\% down to 70 GeV \citep{albert:2008:magic:crab}. For MAGIC-II an energy resolution of $<$25\% for energies down to $\sim50$\,GeV is achievable \citep[see Fig.~3 of][]{colin:2009a}. 
Due to the, on average, higher number of telescopes participating in each event and the increased mirror size of the large telescopes the energy resolution for an NGCTS is expected to improve further. Such an energy resolution paired with an energy spectrum unfolding method \citep[e.g.][]{albert:2007:magic:spectrumunfolding} will enable to robustly reconstruct smooth spectral shapes (e.g. power law or log parabola) even down to low energies. The reconstruction of an EBL attenuation structure at mid-energies is possibly more affected by the limited energy resolution and this will be further discussed at the end of Sect.~4.

\begin{figure}[tbp]
\centering
\includegraphics[width=0.85\textwidth]{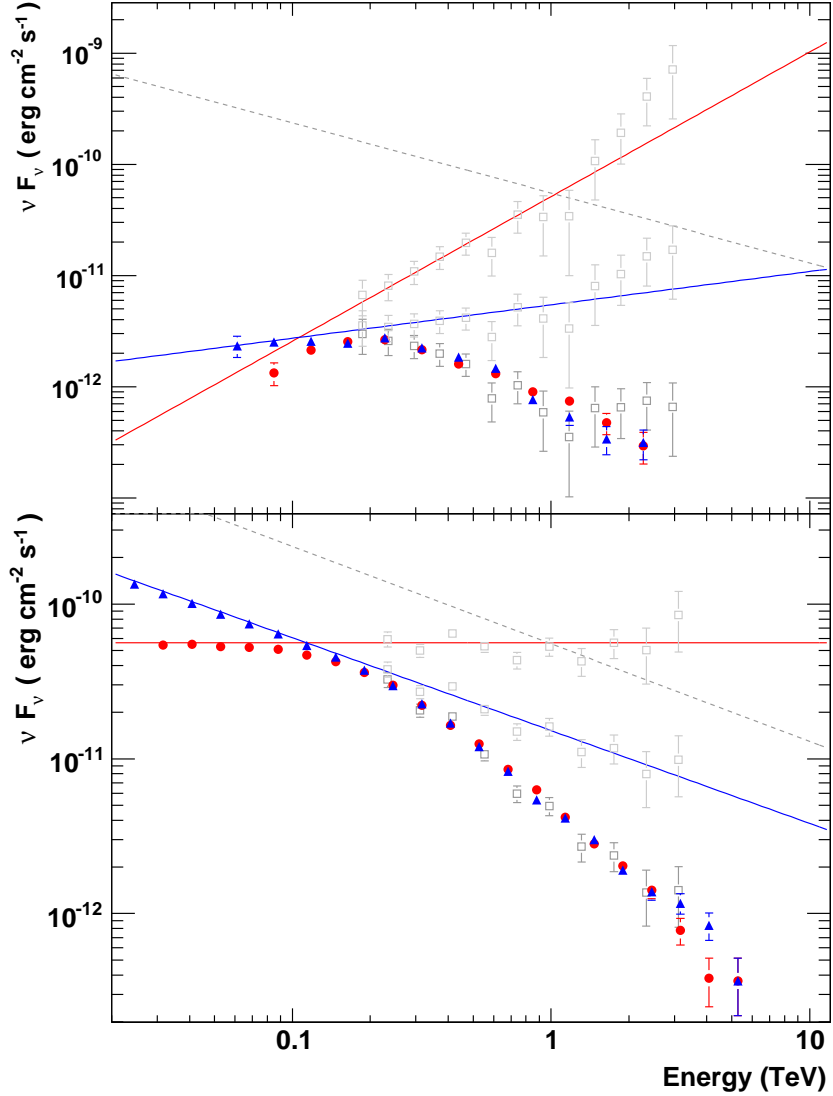}
\caption{Simulated VHE spectra for two sources and two EBL models (blue:Fran08; red:SteckerFE).  \textit{Upper panel:} 1ES\,1101-232. \textit{Lower panel:} PKS\,2155-304. Grey markers show the measured spectra and the measured spectra de-absorbed for EBL attenuation. Red and blue lines are the assumed intrinsic spectra emitted at the source for two different levels of the EBL density, matching the measured de-absorbed spectra. Red and blue markers show the expected spectrum as measured for a next generation VHE instruments simulated in this work. The dashed line gives the flux of the Crab Nebula. Spectral points below 40\,GeV are only shown for illustrative purpose and are not used in the analysis.}
\label{Fig:CTASpecIntrinsic01}
\end{figure}

\paragraph{Simulation example}
Fig.~\ref{Fig:CTASpecIntrinsic01} displays simulated spectra for two sources and two assumed EBL densities for an observation time of 20\,h. The intrinsic spectrum is assumed to follow a simple power law ($\Phi(E)=\Phi_0 \times (E/1\,\mbox{TeV})^{-\Gamma}$), with the parameters adopted so that the EBL attenuated simulated spectrum matches the measured one. Shown are results for 1ES\,1101-232 ($z=0.186$), a hard spectrum ($\Gamma_{VHE} \sim 3$) distant source, whose discovery at VHE energies enabled to derive strong limits on the EBL density in the optical to near-infrared \cite{aharonian:2006:hess:ebl:nature}, and PKS\,2155-304, which is a fairly strong VHE source (about 20\% Crab in the quiescente state) with a softer spectrum ($\Gamma_{VHE} \sim 3.3$) located at an intermediate redshift of $z=0.116$. In the case of 1ES\,1101-232 the hard measured spectrum in combination with the larger distance leads to a very hard intrinsic spectrum (i.e. harder than anticipated by simple leptonic and hadronic models), even for the low EBL model. Such a hard intrinsic spectrum coupled with the relative weakness of the source ($\sim$2\% Crab) leads to a simulated spectrum which does not largely increase the energy range covered compared to the \He measurement of 1ES\,1101-232: neither at low energies (due to the hard spectrum) nor at high energies (the strong EBL attenuation suppresses the signal below the NGCTS sensitivity). Even for the case of the two very different assumed EBL models, no strong difference in the simulated spectra is apparent, which could be used to differentiate between the models. This is different in the case of  PKS\,2155-304, where, for the different EBL models, very different spectra are expected to be measured at lower energies. The behavior at low energies will be discussed further in the next section.

\section{Utilizing the unabsorbed part of the spectrum}\label{unabsorbedspec}

\begin{figure}[tbp]
\centering
\includegraphics[width=0.7\textwidth]{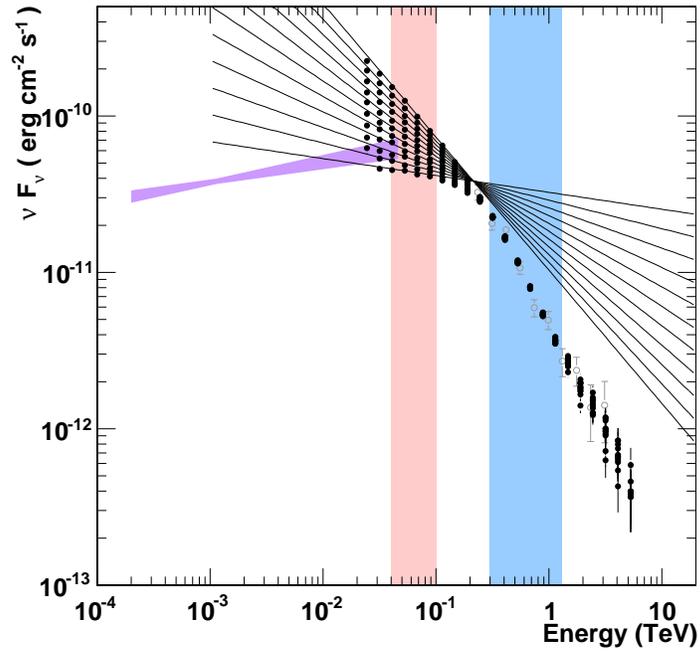}
\caption{Example of simulated spectra for different EBL densities. The base spectrum assumed is the quiescent state spectrum of PKS\,2155-304 ($z=0.116$), EBL model is from Fran08. Shown are: the measured spectrum (grey markers), the simulated spectra for different level of the EBL density (black markers) and the corresponding assumed intrinsic spectra (black lines), the source spectrum in the GeV energy range as measured by Fermi \cite{abdo:2009a:fermi:brightsourcelist:agn} (purple butterfly), and the energy ranges which are used to determine the slope of the simulated spectrum at low (blue) and high (red) energies. Spectral points below 40\,GeV are only shown for illustrative purpose and are not used in the analysis.}
\label{Fig:LowEnergiesExample}
\end{figure}
 
\begin{figure}[tbp]
\centering
\includegraphics[width=0.75\textwidth]{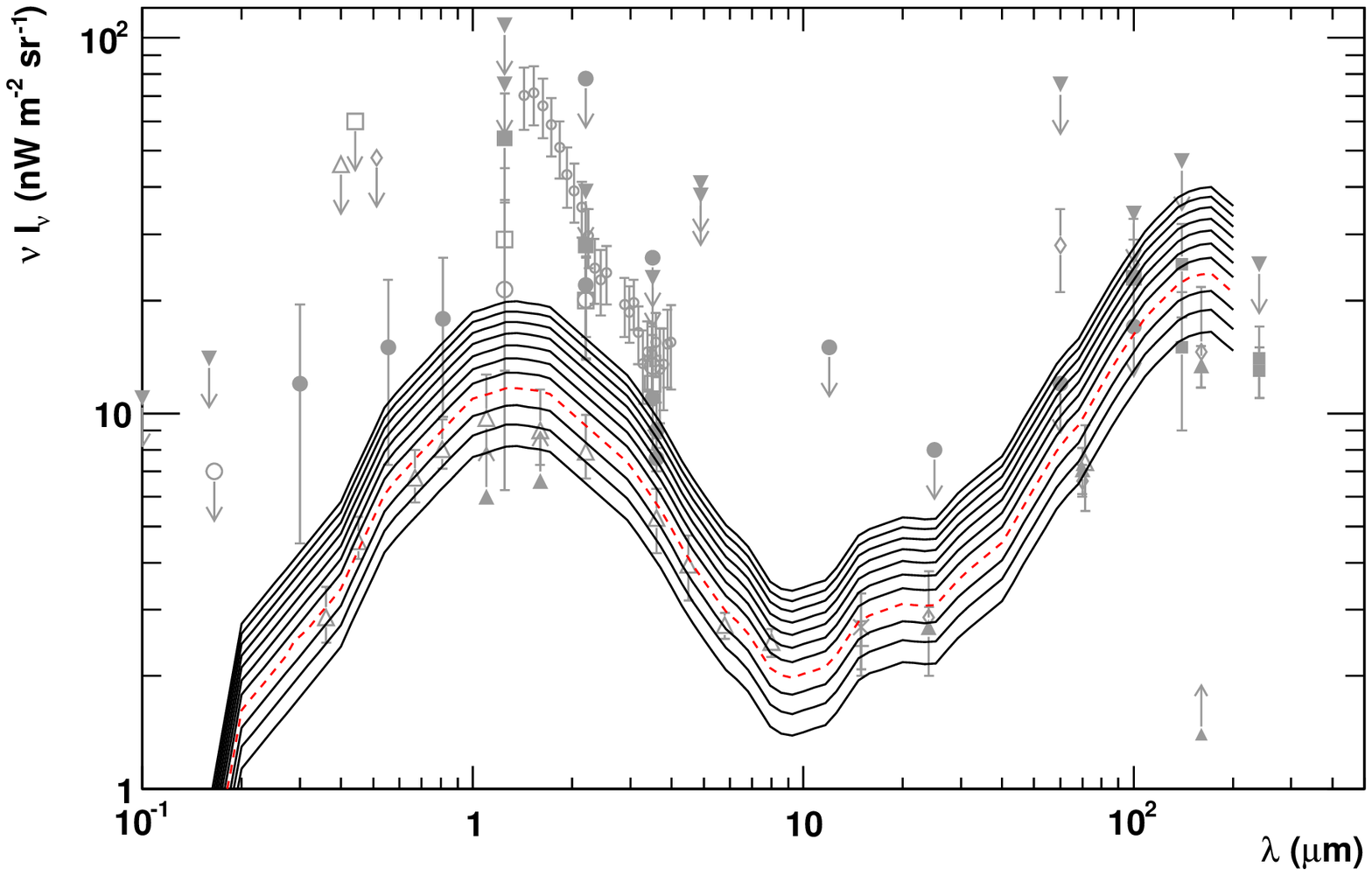}
\caption{EBL density ($z=0$) for the Fran08 EBL model scaled with a factor 0.7 to 1.7 in steps of 0.1. The red dashed curve is the unscaled EBL model.}
\label{Fig:EBLModelFran08Scaled}
\end{figure}

One of the main features of a NGCTS will be a high sensitivity in the energy range between 20 and 100\,GeV, an energy range which holds the possibility to directly sample parts of the energy spectrum of a source, which are not affected by the EBL attenuation. In this section it will be explored how this energy range can be utilized to derive limits on the EBL density, and what are possible problems and caveats.

\begin{figure}[tbp]
\centering
\includegraphics[width=0.33\textwidth]{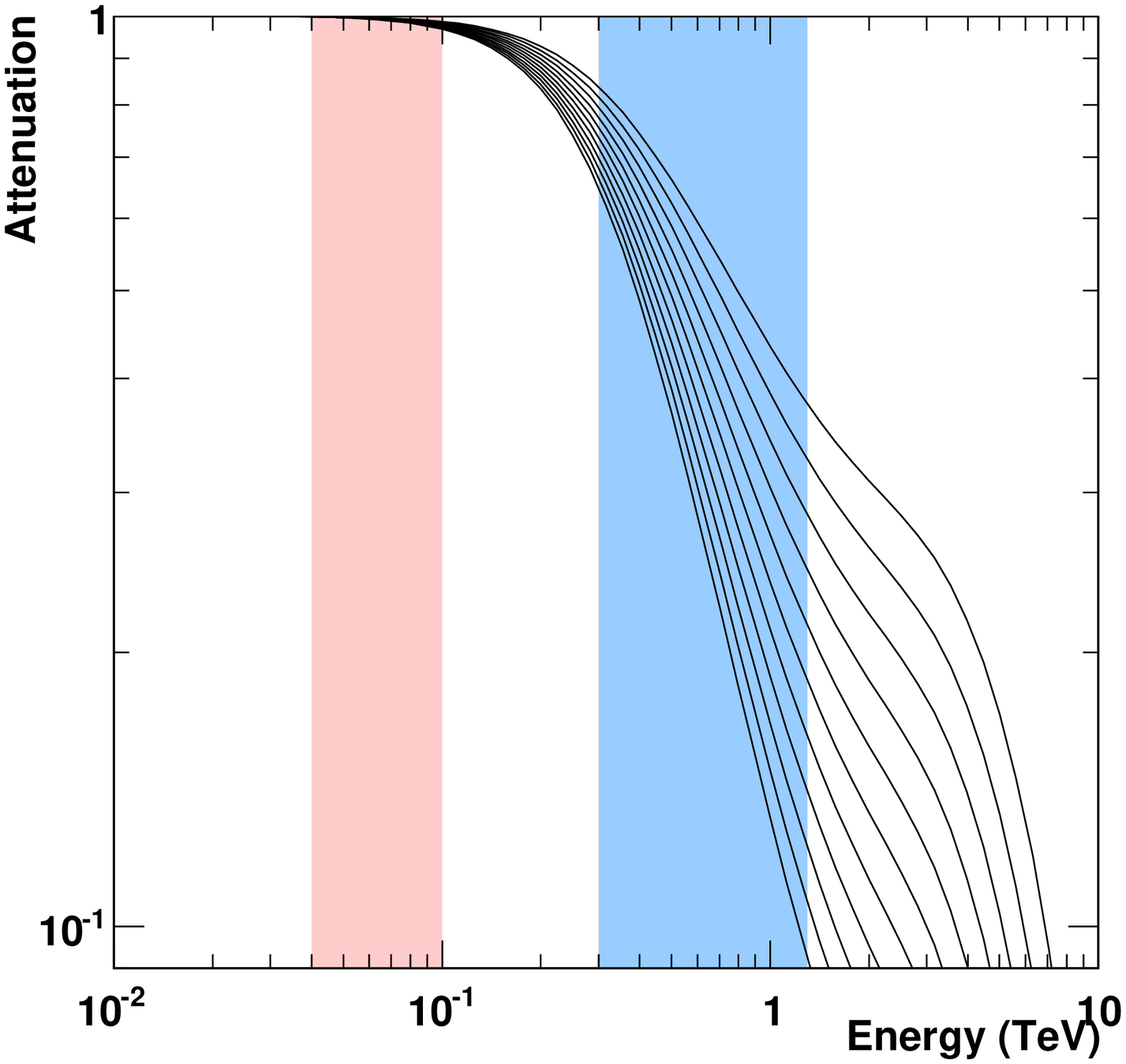}\hfill%
\includegraphics[width=0.33\textwidth]{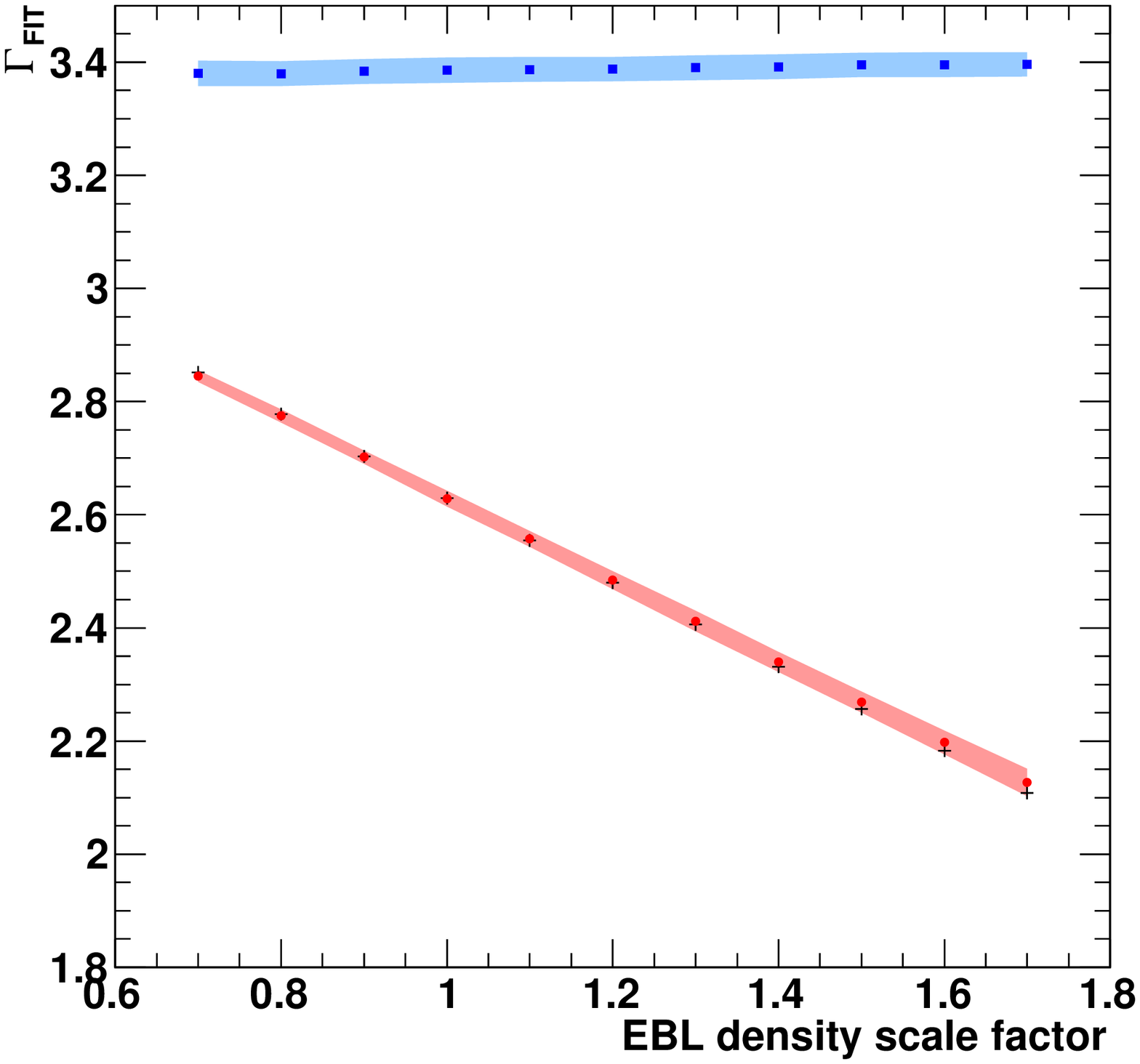}\hfill%
\includegraphics[width=0.33\textwidth]{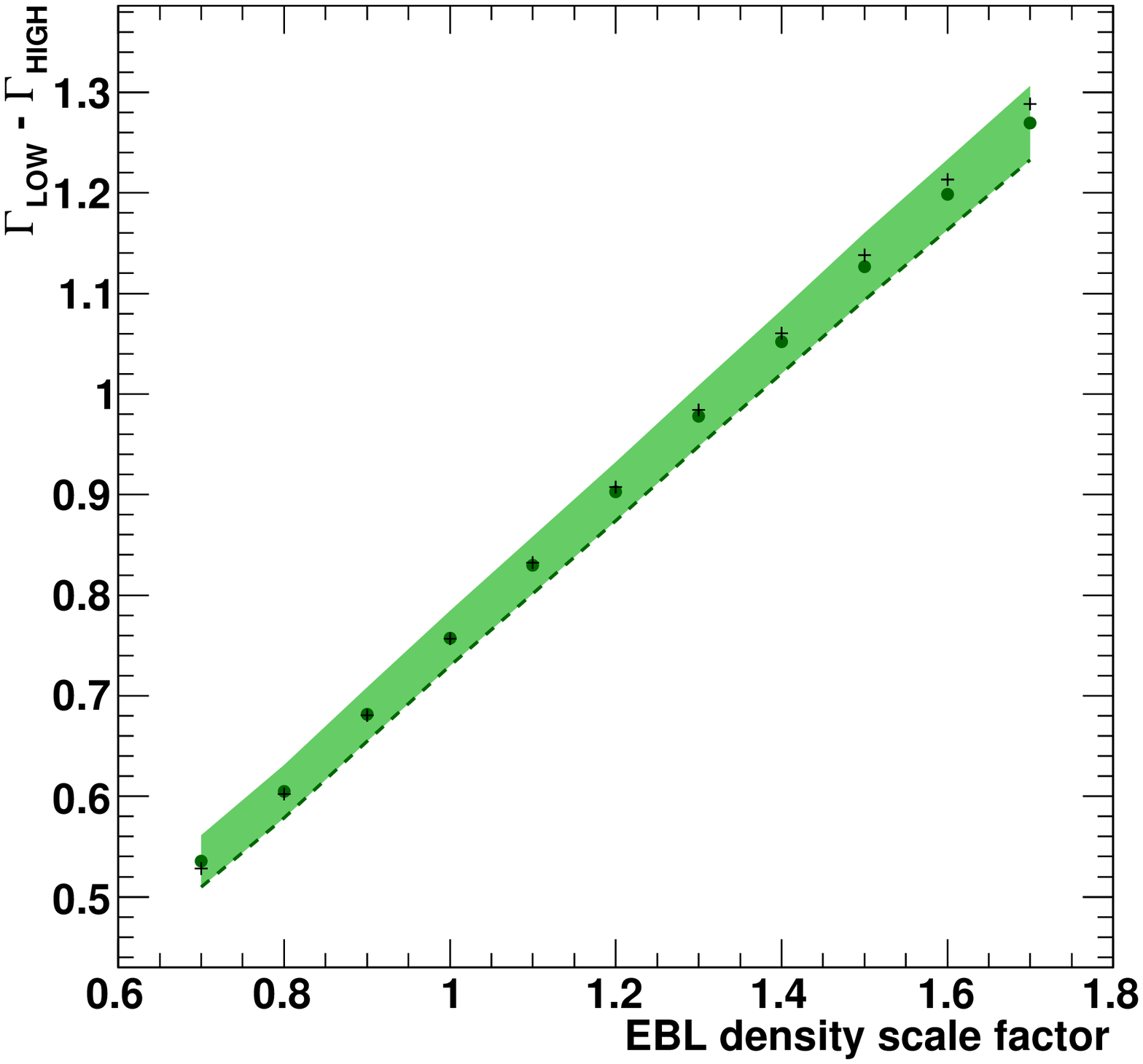}\\
\includegraphics[width=0.33\textwidth]{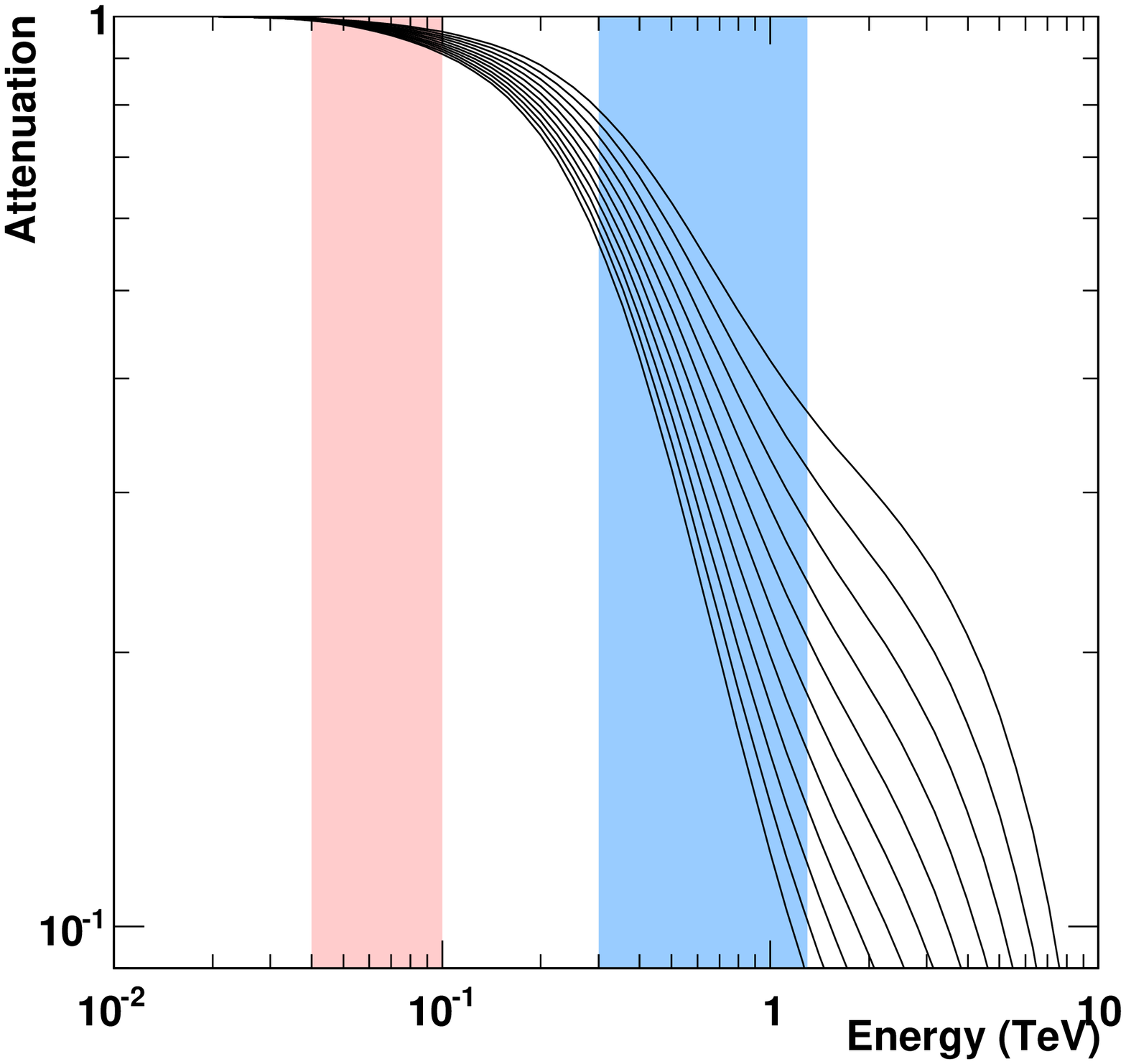}\hfill%
\includegraphics[width=0.33\textwidth]{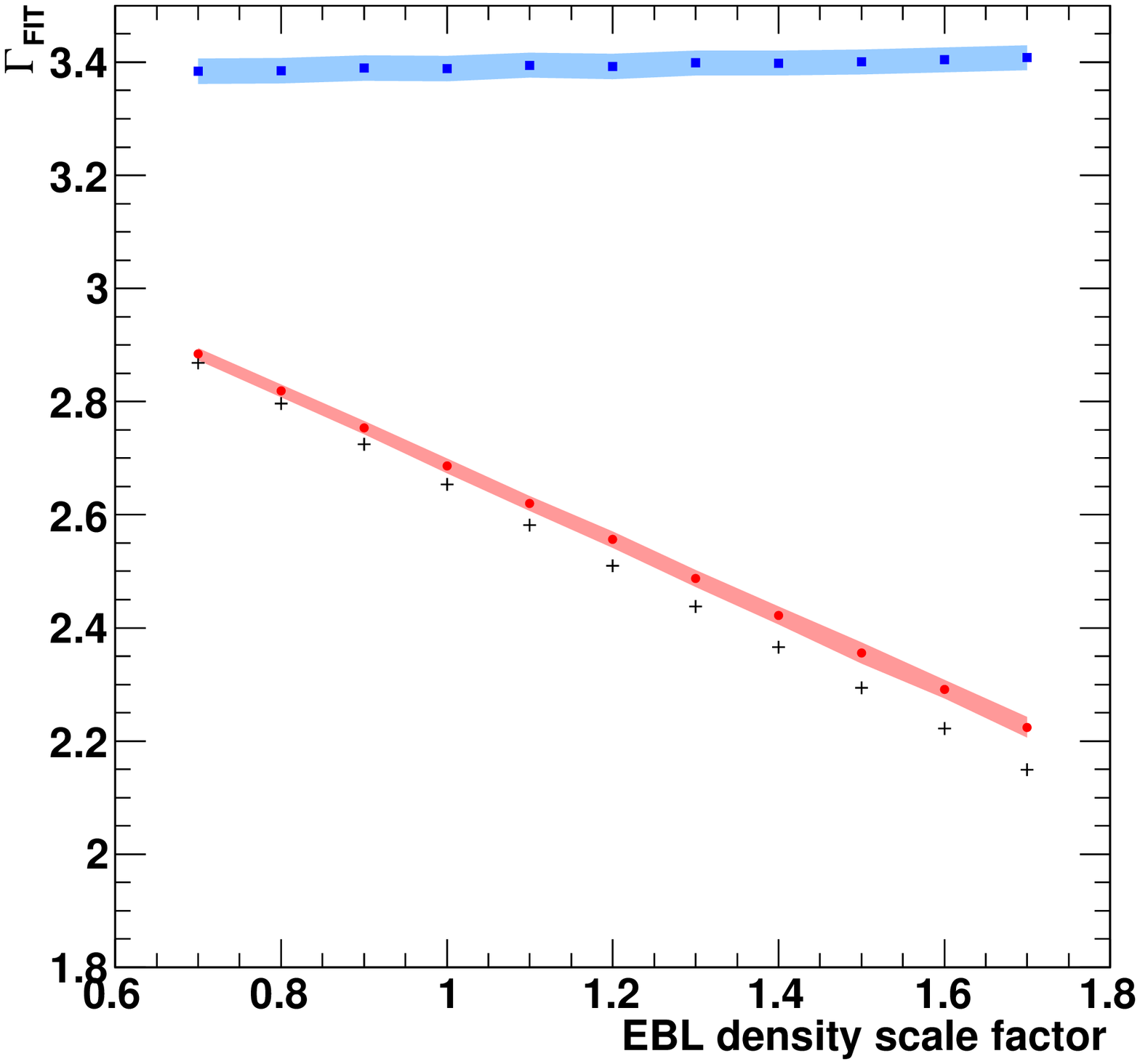}\hfill%
\includegraphics[width=0.33\textwidth]{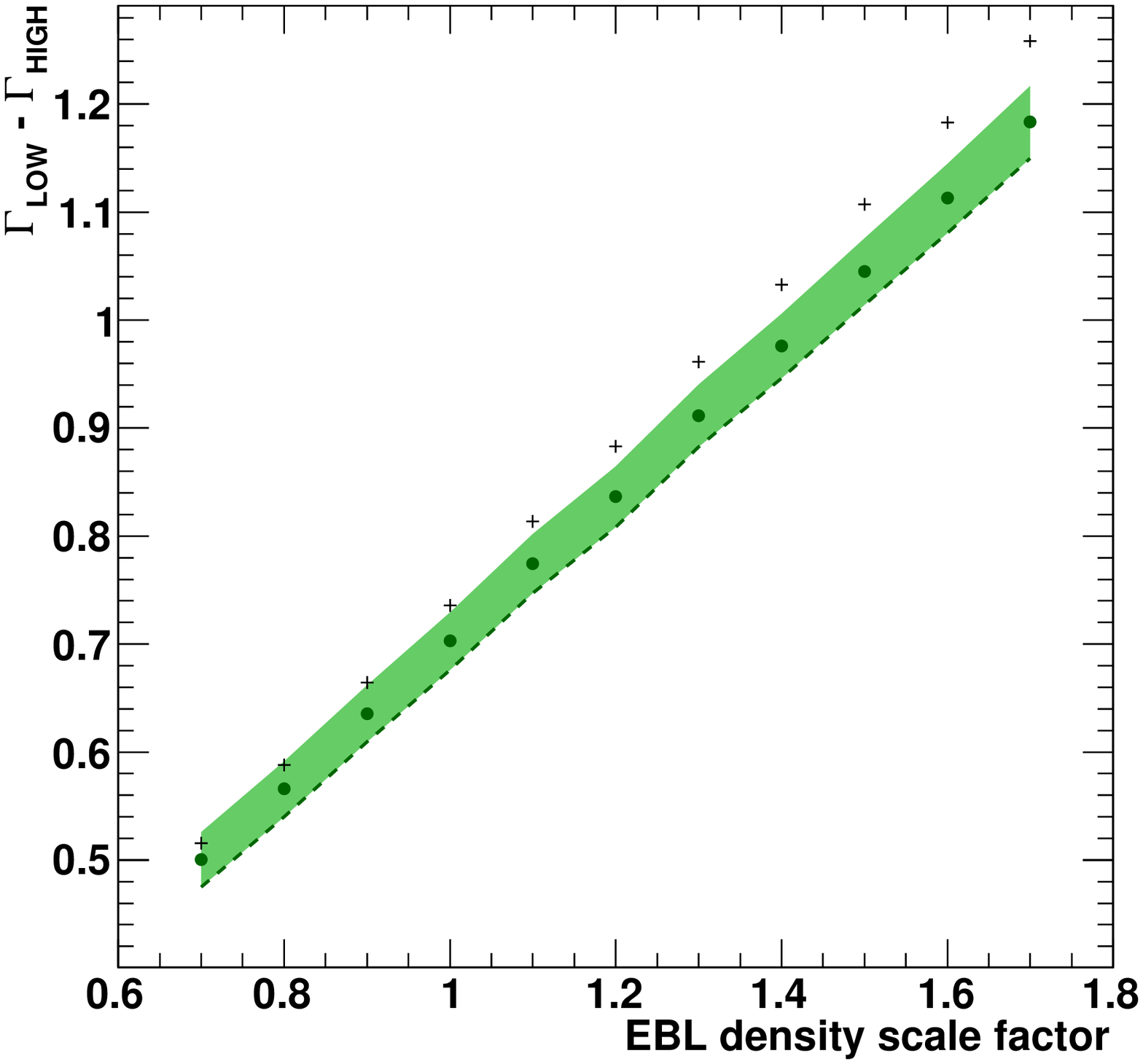}
\caption{Results from the power law fit to the simulated spectrum in the low and high energy band using the PKS\,2155-304 quiescence spectrum with the Fran08 (upper row) and Gen (lower row) EBL model. \textit{Left:} Attenuation of the VHE $\gamma$-ray flux resulting from the scaled EBL models. The colored boxes in the background mark the fit range for the power law at low (red) and high (blue) energies. \textit{Middle:} Spectral index $\Gamma$ from the fit of a power law to the low (red) and high (blue) energy range versus the scaling factor of the EBL model. The color shaded bands denote the error on the spectral index from the fit (RMS of the spectral index distribution). Black crosses mark the intrinsic spectral index that has been utilized. \textit{Right:} Spectral break between low and high energies (error bands from error propagation). For discussion see main text.}
\label{Fig:LowEnergiesPKS2155PLResults}
\end{figure}
 
\paragraph{Simulation \& analysis chain}
The following simulation and analysis chain is utilized:
\begin{enumerate}
\item Calculate EBL attenuation for a specific EBL density and source distance.
\item De-attenuate a measured spectrum and fit with power law ($dN/dE = \Phi_0 \cdot  E^{-\Gamma}$). The fit results serve as input flux function for the simulated spectrum.
\item Simulate spectrum as measured by an NGCTS with calculated input flux function.
\item Fit simulated spectrum in a low energy regime (intrinsic spectrum) and high energy regime (absorbed spectrum). Again, a simple power law function is used in each energy regime.
\item 200 spectra are simulated and mean values are used.
\end{enumerate}
An example for this procedure is shown in Fig.~\ref{Fig:LowEnergiesExample} for the quiescence spectrum of PKS\,2155-304 ($z=0.116$). An observation time of 20\,h has been assumed, which could easily be extended given that this is a steady flux state. As EBL model the Fran08 model is used, scaled in steps of 0.1 from 0.7 to 1.7 (Fig.~\ref{Fig:EBLModelFran08Scaled}).

\paragraph{Fit-range \& source intrinsic break}
The resulting EBL attenuations for the scaled models are shown in Fig.~\ref{Fig:LowEnergiesPKS2155PLResults} upper left panel. The energy ranges used to fit the power laws are marked by colored boxes. In the high energy range the attenuation follows approximately a power law and since the input spectrum is also a power law, the resulting attenuated spectrum will follow a power law as well. In the low energy range no significant absorption is present. As recent \Fe/LAT observations show, a spectral break is observed between the GeV and the TeV energy range \cite{abdo:2009:fermi:tevselectedagn}. For many sources this break can be attributed to a break expected from EBL attenuation. For these sources an NGCTS would be able to sample the intrinsic spectrum down to very low energies. In other cases, e.g. PKS\,2155-304, the break between GeV and TeV range is stronger than expected from EBL attenuation, therefore an additional, source intrinsic break is expected somewhere between the two energy bands. Currently, the statistics for this energy range for the sources considered here is not yet sufficient to correctly model the break. Therefore, for the analysis in this paper the lower edge of the low energy fit range is chosen to start above the \Fe/LAT energy band (defined as the 5th highest photon in energy reported in \cite{abdo:2009a:fermi:brightsourcelist:agn}, e.g. $\sim$40\,GeV for PKS\,2155-304). Future observations with \Fe/LAT, \MaII, and \HeII will provide further information on this issue.

\paragraph{EBL attenuation spectral break}
Fig.~\ref{Fig:LowEnergiesPKS2155PLResults} upper middle panel shows the spectral index resulting from the power law fit for the two energy bands for the different scalings of the EBL model. The markers show the mean spectral index of the fits, with the error given as the RMS of the mean spectral index distribution (shaded bands). As alternative error definition the mean error on the spectral index from the fit could be used which gives similar results. The black crosses mark the spectral index utilized for the input source spectrum. It can be seen that the fit in the low energy band reproduces very well the assumed intrinsic spectral indices.\footnote{The correlation between the EBL density and the spectral index in the low energy band is a result of the methodology i.e. for each EBL density a corresponding intrinsic spectrum is constructed.} For the highest scaled EBL densities the effect of attenuation becomes relevant in the low energy band and the spectral index from the fit is steeper (larger) than the one from the input spectrum. This effect will be discussed in more detail below.
The right panel of the figures displays the strength of the spectral break between the two power laws $\Gamma_{LOW} - \Gamma_{HIGH}$, with the error bands calculated via error propagation. An EBL density scale factor of 0.1 corresponds to one to two standard deviations difference in the strength of the break. \footnote{I.e. if the Fran08 EBL model is correct and the intrinsic spectrum follows a power law in the energy range considered a 1.3 scaling of the model could be excluded with 3 standard deviations.} Note that a scaling of 0.1 corresponds to an EBL density of $\sim$1\,nW\,m$^{-2}$sr$^{-1}$ at 2\,$\mu$m which is of the same order as the error on the lower limits from integrated source counts at this wavelength. The detection of a break with certain strength can be converted into an upper limit of the EBL density (thick dashed line). In principal, a single well measured spectrum is sufficient to derive such limits. Since EBL attenuation is a global, redshift dependent phenomenon, in general, a combined fit to all available VHE data should be used to derive limits on the EBL density \citep[see e.g.][]{mazin:2007a}. This is discussed in more details in Sect.~\ref{conclusions}.

\paragraph{EBL attenuation at low energies}
The same analysis has been performed utilizing the Gen EBL model and the results are displayed in Fig.~\ref{Fig:LowEnergiesPKS2155PLResults} lower row. The Gen EBL density in the ultraviolet to optical is higher than in the Fran08 model (Fig.~\ref{Fig:EBLModels}), therefore resulting in a non negligible attenuation in the low energy band (Fig.~\ref{Fig:LowEnergiesPKS2155PLResults} lower left panel). Consequently, the power law fit in the low energy band results in a too soft spectrum compared to the input source spectrum (Fig.~\ref{Fig:LowEnergiesPKS2155PLResults} lower middle panel). This mis-reconstruction, if taken at face value, would lead to an overestimated upper limit on the EBL density (Fig.~\ref{Fig:LowEnergiesPKS2155PLResults} lower right panel). To be able to utilize the low energy part of the spectrum as proxy for the intrinsic spectrum it is therefore crucial to carefully examine the spectral shape in the energy range for signs of curvature. A conservative approach would be to correct the flux points in the fit energy range for the attenuation from a maximum EBL before performing the fit. On the other hand, if a high statistic VHE spectrum from a distant source does not show any indication of curvature in this energy regime, tight limits on the EBL density in the ultraviolet to optical EBL can be derived.

\begin{figure}[tbp]
\centering
\includegraphics[width=0.5\textwidth]{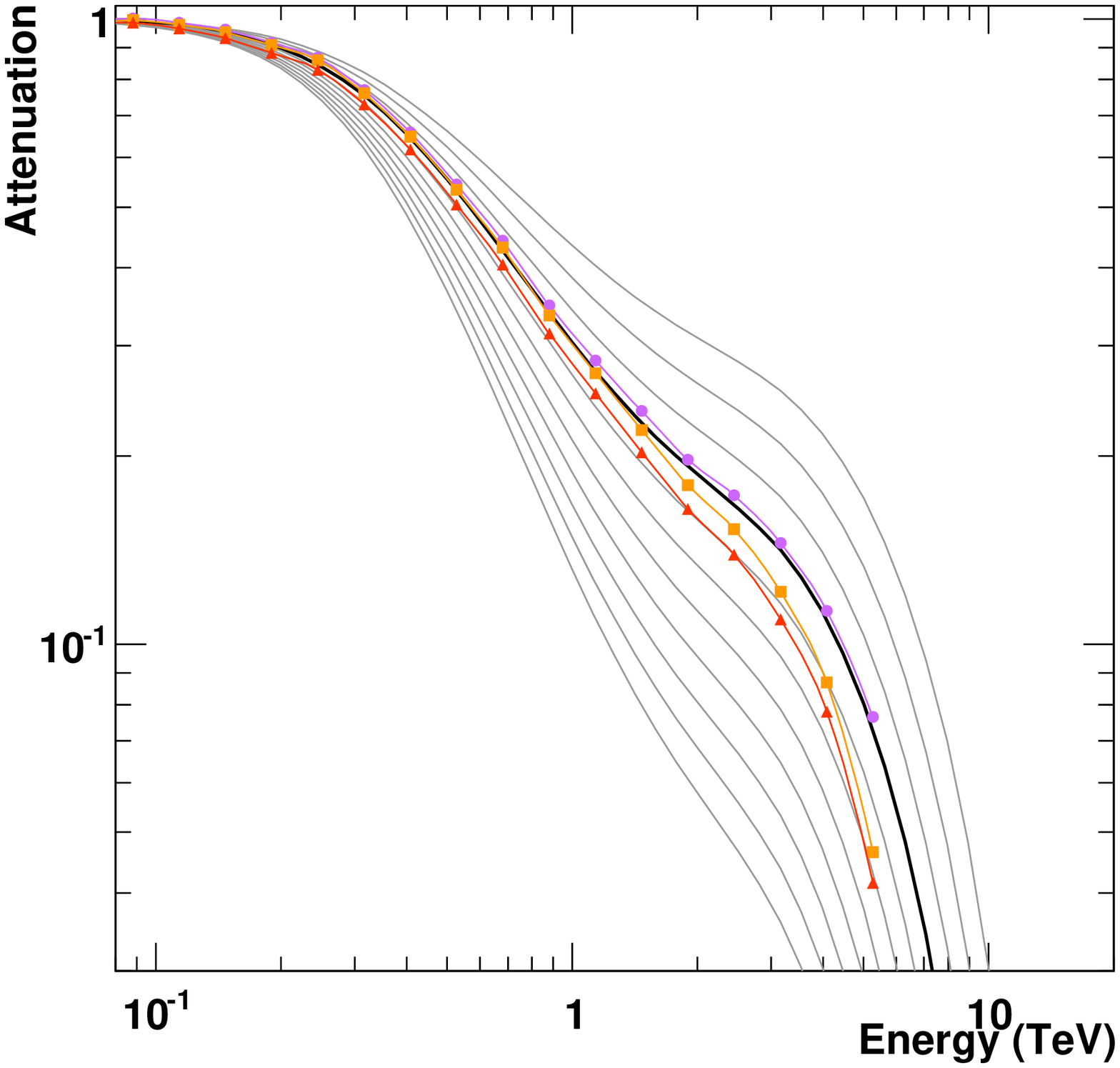}\hfill%
\includegraphics[width=0.5\textwidth]{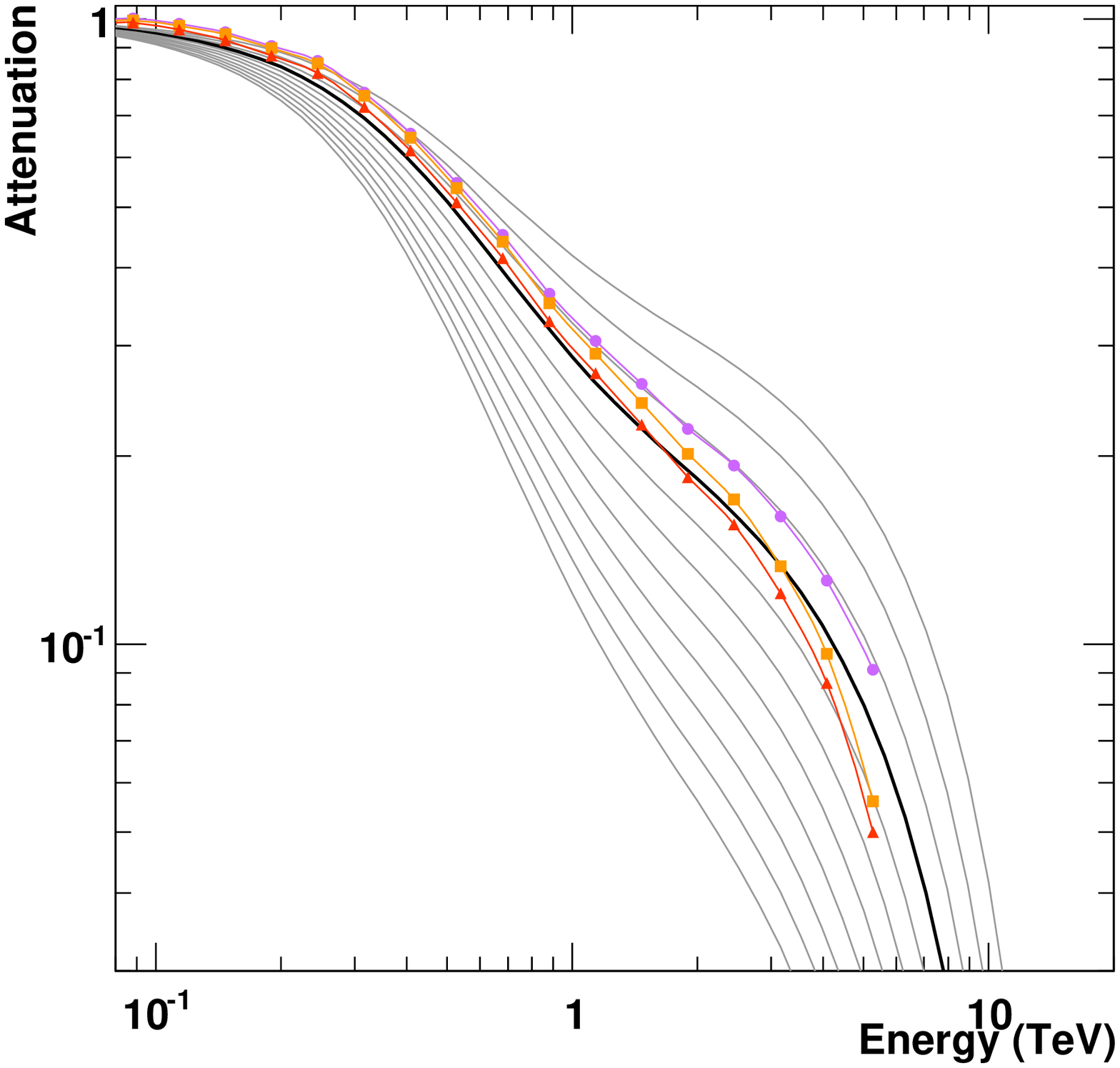}
\caption{Energy dependent upper limits on the attenuation for the Fran08 (left) and the Gen (right) EBL models derived utilizing the PKS\,2155-304 quiescence spectrum. Shown are the limits taking into account the 1$\sigma$ error on the flux points (orange filled squares), with the addition of the 1$\sigma$ error on the spectral index of the low energy power law (red filled triangles), and no errors (purple filled circles). The black line shows the attenuation for the EBL model utilized, grey lines show the attenuation for the EBL model scaled in steps of 0.1.}
\label{Fig:LowEnergiesPKS2155AttResults}
\end{figure}
 
\paragraph{Energy dependent attenuation limits}
The power law fit to the low energy region, if taken as proxy for the intrinsic spectrum, can also be used to derive energy dependent upper limits on the attenuation. Such limits can be extremely useful: the EBL attenuation is an energy dependent process, where certain wavelength regions of the EBL spectrum are connected to certain VHE energy ranges. Features in the EBL density, as e.g. expected from the first stars \cite[see e.g.][]{raue:2009a}, could therefore also produce features in the attenuation. The resulting energy dependent EBL attenuation limits from such an analysis for the Fran08 and the Gen EBL model utilizing the PKS\,2155-304 quiesent spectrum are shown in Fig.~\ref{Fig:LowEnergiesPKS2155AttResults}\footnote{For this investigation only a single simulated spectrum without applying the statistical dicing of the event numbers is used.}. As can be seen from the figure, the dominant source of errors in such an analysis are the uncertainties on the spectral index from the power law fit in the low energy range, since the statistical errors on individual flux points are comparably small. Again, the rejection power is in the order of 1-2\,$\sigma$ per 0.1 scaling of the EBL density at energies $>1$\,TeV, when utilizing a simple and ad-hoc error definition. In the case of the Gen EBL model again the effect of the attenuation "spill over" in the low energy fitting band is visible: due to the EBL attenuation the power law fit is too soft, resulting in an underestimated EBL attenuation and too strong limits.

\begin{figure}[tbp]
\centering
\includegraphics[width=0.5\textwidth]{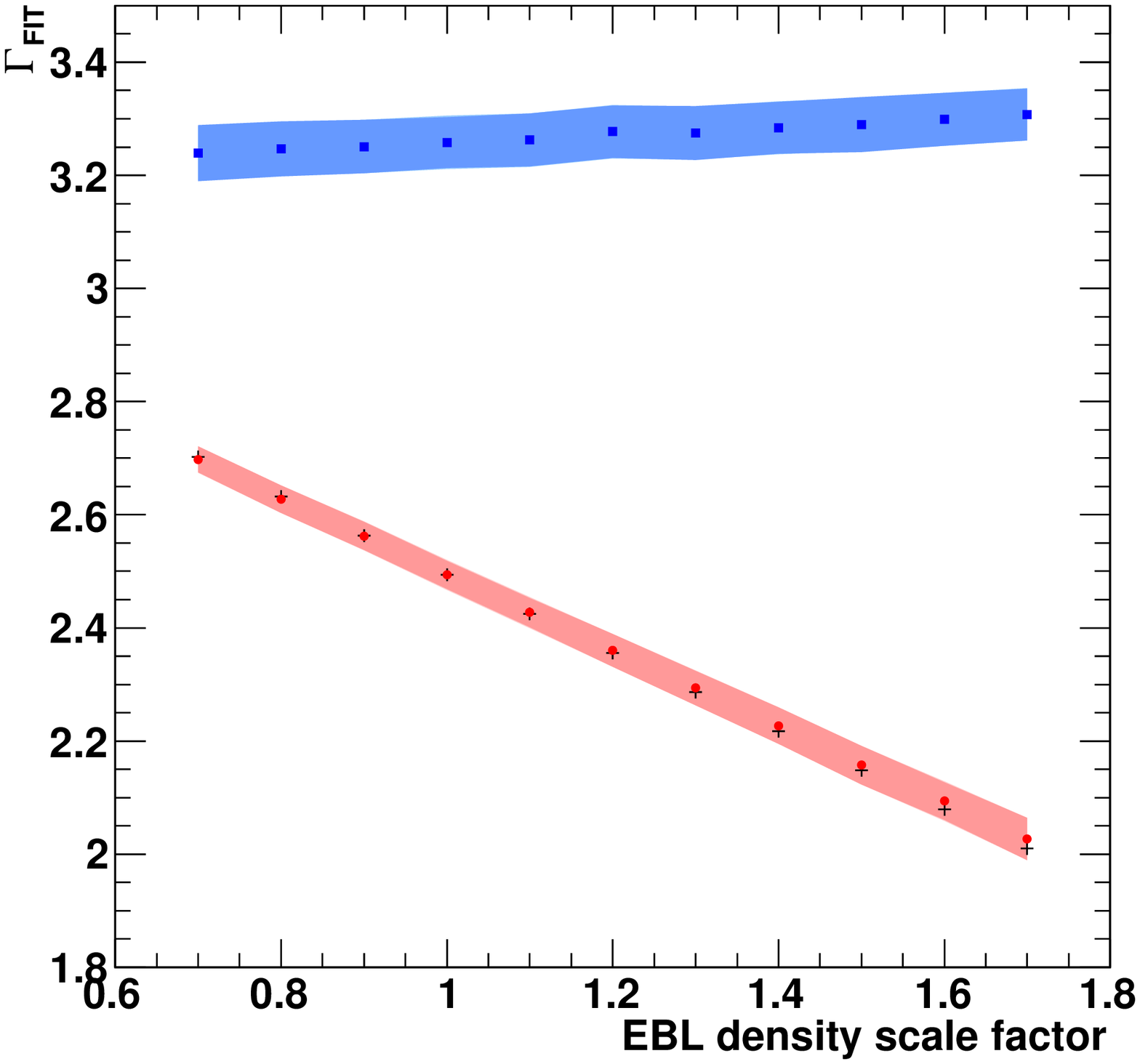}\hfill%
\includegraphics[width=0.5\textwidth]{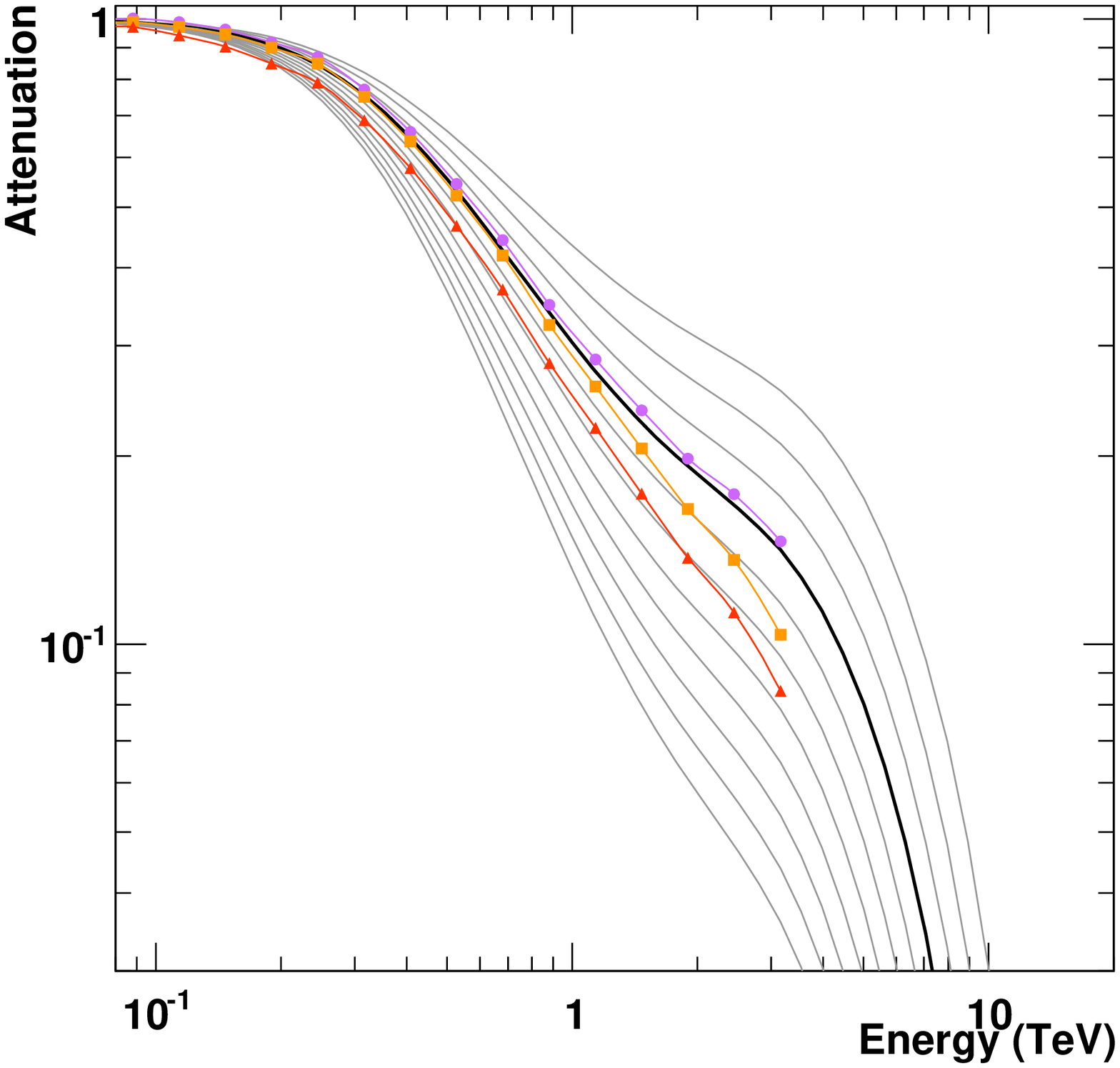}\\
\includegraphics[width=0.5\textwidth]{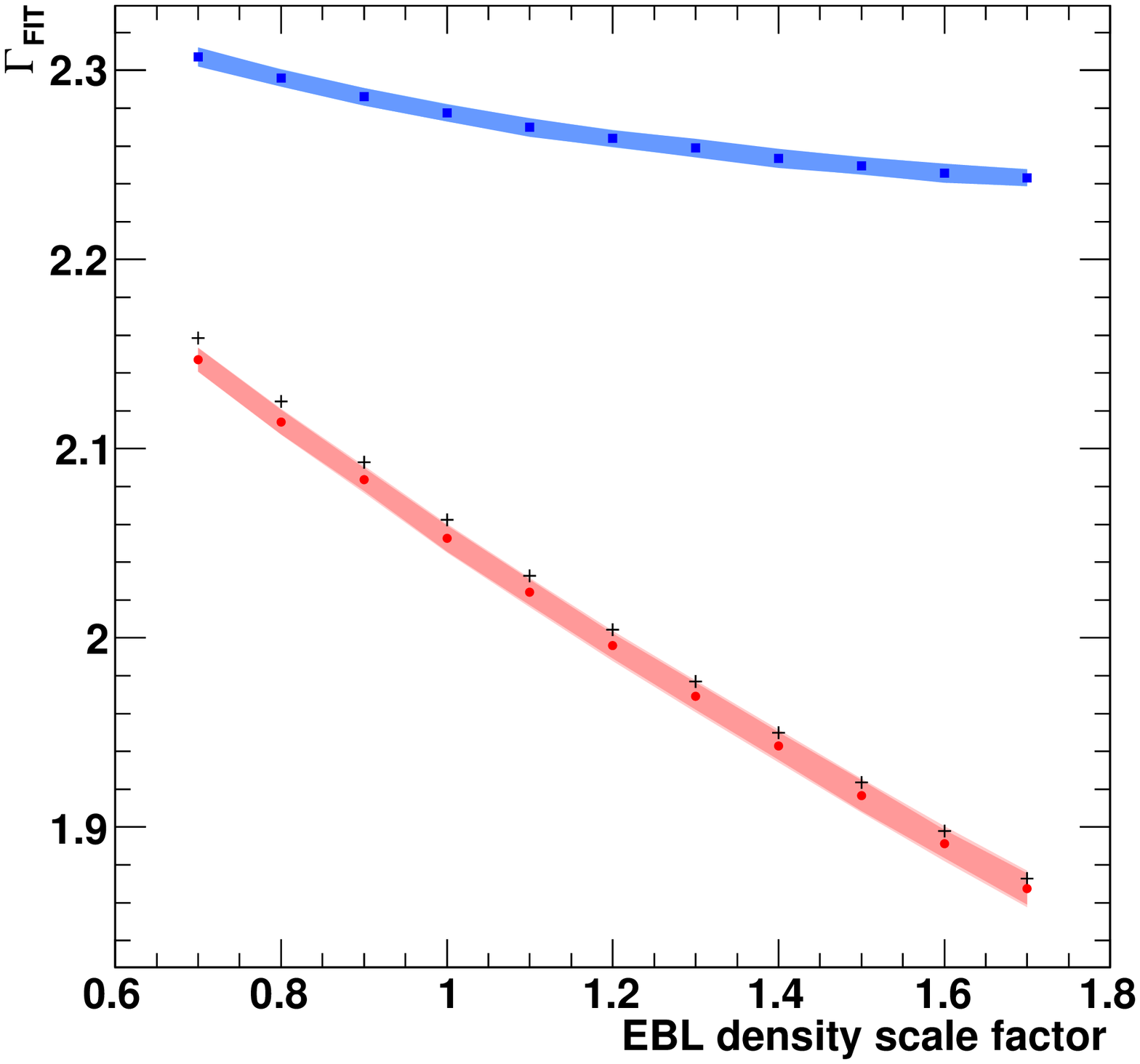}\hfill%
\includegraphics[width=0.5\textwidth]{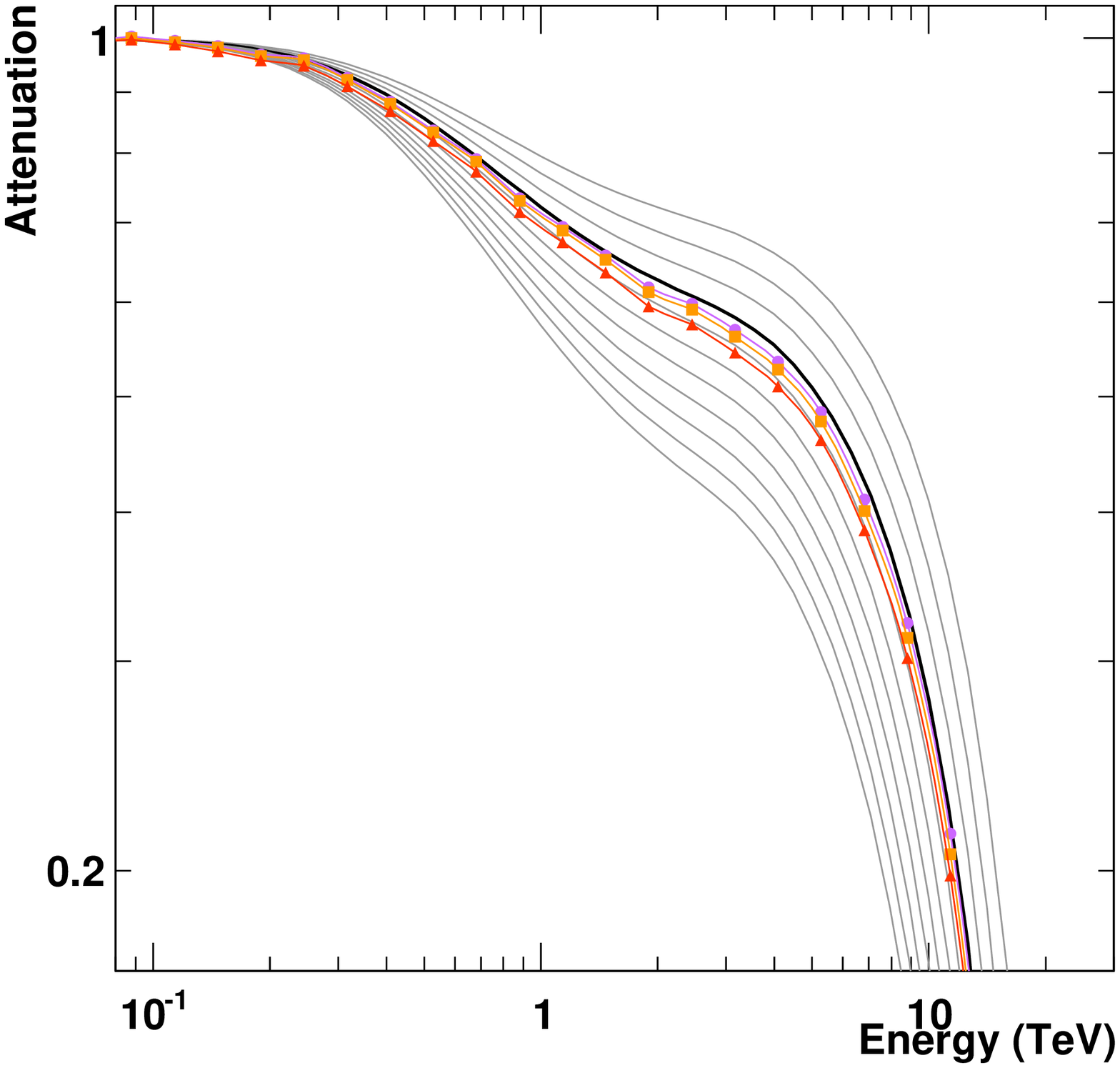}
\caption{EBL limit results from power law fits to low and high energy bands of the VHE spectrum  using scaled version of the Fran08 EBL model (see Fig.~\ref{Fig:LowEnergiesPKS2155PLResults} and \ref{Fig:LowEnergiesPKS2155AttResults} for detailed description). \textit{Upper row:} Results for the VHE spectrum from the flaring state of PKS\,2155-304 in 2006 \cite{aharonian:2007:hess:pks2155:bigflare}  (6\,min observation time), \textit{Lower row:} Results for the averaged high state spectrum of Mkn\,501 \cite{aharonian:1999b} ($z = 0.034$; 20\,h observation time).}
\label{Fig:LowEnergiesMisc1}
\end{figure}

\paragraph{Flaring states}
Up to here only the quiescent flux state of sources has been considered, since its observation is guaranteed. AGNs do also show extreme flaring behavior with high fluxes, unfortunately usually only for short periods of time. Here, the VHE spectrum recorded during the extreme flux outburst of PKS\,2155-304 in 2006 \cite{aharonian:2007:hess:pks2155:bigflare} is investigated. The spectrum is an averaged spectrum of 1.5\,h of observations, in which the source showed several strong flares  with time-scales down to a few minutes. While no strong spectral variations in the VHE spectrum have been observed during this flare, for the analysis only short observations of 6\,min length have been considered, demonstrating the power of the NGCTS to derive high quality spectra even on such short time-scales. In addition, the EBL attenuation is a "stationary" effect in the spectrum, therefore a temporally fine resolved flaring state might enable to distinguish between intrinsic and EBL effects. The resolution of the EBL density achieved for such a 6\,min observation of the source in a flaring state is comparable with the one achieved for 20\,h of quiescence observations (Fig.~\ref{Fig:LowEnergiesMisc1} top row).

\paragraph{Nearby sources}
Strong flux outbursts in the VHE regime have also been observed in the two nearby sources Mkn\,421 and Mkn\,501. While the smaller distances result in less EBL attenuation and therefore a weaker signature, the overall higher flux levels of these two sources enable to derive spectra with high event statistics. Here, the time-averaged VHE spectrum of Mkn\,501 during the 1997 high state is investigated, assuming an observation time of 20\,h. As can be seen from Fig.~\ref{Fig:LowEnergiesMisc1} bottom row, while the overall strength of the break between the two power laws is much smaller than in the case of PKS\,2155-304 ($0.15-0.4$ vs $0.5-1.2$) comparable limits can be derived due to the superior statistics. In addition, the problem of attenuation in the low energy band is less severe. One caveat - at least for the case of Mkn\,421 - is the fact that there are strong indications for an intrinsic (not EBL related) break between the GeV and TeV range at relatively high energies (around 100\,GeV) \cite{abdo:2009:fermi:tevselectedagn}, which would make it difficult to define a proper low energy region void of any curvature. A different method, discussing the possibilities of how such high quality spectra with a wide energy range produced by a nearby source can be used to test the EBL density, is discussed in the next section.

\section{Attenuation modulation at mid-energies}

The NGCTS will achieve an unprecedented sensitivity in the intermediate energy range, i.e., between 100\,GeV and few TeV. For nearby sources, this energy range is most sensitive for the EBL density in the optical to infrared regime
 ($\sim$1-15$\,\mu$m), which can, thus, be probed very efficiently with an NGCTS measurement of an AGN energy spectrum.

\paragraph{Smoothness of AGN spectra}
The measured energy spectra of AGNs in the energy range between 100\,GeV and few TeV follow usually a smooth shape. For most of the measured sources, a simple power law fit is sufficient to describe the available data well, whereas for sources in a flare state (like the flare of PKS\,2155-304 in 2006) or with a generally high emission state (like Mkn\,421), either a curved power law  or a power law with a cut-off are successfully used. The curved power law (also known in the literature as the double-log parabola) is expected to describe the spectra well at energies 
close to the position of the Inverse-Compton peak. The power law with a cut-off instead is the expected behavior of a source  which does not provide necessary conditions for acceleration of charged particles to sufficiently high energies. All scenarios do have one common feature: the measured spectra can be described by smooth functions,i.e., no features, wiggles or pile-ups are expected, especially after de-convolving the spectra for the effect of the EBL absorption.
This property can, therefore, be used to distinguish between different overall EBL levels in the optical to infrared regime: whereas the "correct" EBL model and level will produce a smooth intrinsic AGN spectrum, an "incorrect" EBL level would result in a signature (in form of well defined wiggles) in the reconstructed intrinsic spectrum.

The strength of the method is that the EBL signatures in the reconstructed AGN spectra will
not only be visible (measurable) in the case where the assumed EBL level is higher than the real one, but also in case the assumed EBL level is lower than the real one. It is, therefore, the first
indirect method to really measure the EBL density at z=0.

\paragraph{Simulation \& analysis chain}
The utilized methodology is sketched below:
\begin{enumerate}
 \item Assume an intrinsic spectral shape and the flux level of a known strong extragalactic gamma-ray source.
 \item Simulate NGCTS spectrum assuming the absorption due to the standard EBL, i.e., with the scaling factor of 1.
 \item Reconstruct intrinsic spectrum of the source assuming a scaled EBL level. For a scaling, which is different enough from the standard EBL, the reconstructed AGN spectrum will
show distinct wiggles.
 \item To characterize the presence of the wiggles, a fit by a smooth source function (spectral shape) is performed.  
The chosen fit shape is the curved power law: $\mathrm{d}N / \mathrm{d}E = N_{0} \times \left( E^{-\alpha + \beta\,\log(E/E_{0})} \right)$ 
 \item The resulting $\chi^{2}$ of the fit is then used to judge if the change in the EBL level results in an improbable reconstructed intrinsic spectrum. 
 \item Repeat the simulation 1000 times for the given EBL scaling and compute the mean and the RMS of $\chi^{2}$  values from the fits to the reconstructed intrinsic 
spectrum.
\end{enumerate}

\begin{figure}[tbp]
\centering
\includegraphics[width=0.75\textwidth]{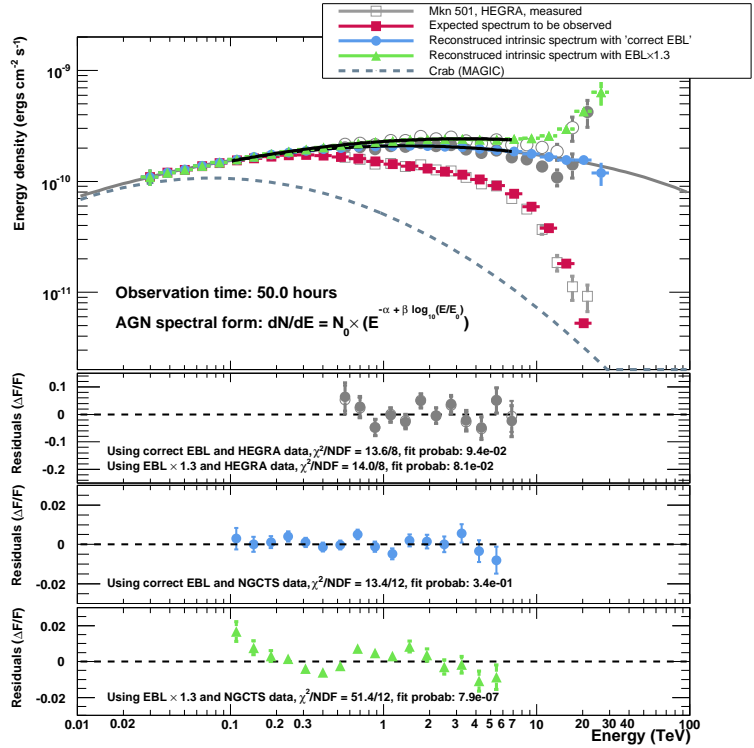}
\caption{
Search for signatures at mid-VHE in the Mkn\,501 spectrum.
\textit{Top panel:} The Mkn\,501 spectral energy distribution (SED) at very high
energies. Shown are: measured spectrum by HEGRA (grey open squares),
simulated NGCTS spectrum (red filled squares),
assumed intrinsic spectrum of Mkn\,501 (grey solid line),
reconstructed intrinsic Mkn\,501 spectrum (blue filled circles) and reconstructed
intrinsic spectrum of Mkn\,501 assuming an EBL scaling factor of 1.3 (green triangles).
For comparison, the deabsorbed HEGRA Mkn\,501 is shown for the same two EBL models (filled grey circles and open grey circles, respectively). The Crab Nebula spectrum (grey dashed line) is shown for comparison.
\textit{Middle upper panel:} Residuals of the fits to the de-absorbed HEGRA Mrk\,501 spectrum for the case of the two different assumptions about the EBL density. The difference between the two curves (filled and open circles) is very small. The HEGRA sensitivity is not sufficient to decide between the two assumptions.
\textit{Middle lower panel:} Residuals between the best fit to the SED in case of NGCTS measurement and the spectral points from the intrinsic spectrum using the correct EBL density.
\textit{Bottom panel:} Residuals between the best fit to the SED for a NGCTS  measurement and the spectral points from the intrinsic spectrum reconstructed using the scaled EBL model.
A clear and significant signature (wiggles) in the residuals is visible, which is quantified by
a low probability of the fit.}
\label{Fig:MidEnergyExampleOfMethod}
\end{figure}
 
\paragraph{Simulation example}
The steps 1--4 of the method are illustrated in
Fig.~\ref{Fig:MidEnergyExampleOfMethod} for the VHE spectrum of Mkn\,501. The assumed
spectral shape and the flux level of the intrinsic spectrum are adapted to the
flux measured by HEGRA  \cite{aharonian:1999b} during the outburst of the source in 1997: the
original data are shown by grey open squares in the upper panel of the figure.
The simulated NGCTS spectrum calculated using the 'correct EBL' (Gen EBL model; scale factor 1)
is shown in red, whereas the assumed intrinsic spectrum of Mkn\,501
is shown by the solid grey line and the reconstructed intrinsic spectrum
is shown by the blue filled circles.
The effect of the mis-reconstruction of the intrinsic spectrum is shown for the example of
an EBL scaled by a factor of 1.3: the reconstructed intrinsic spectrum (green filled triangles)
clearly shows wiggles in the fit range. The effect of the wiggles is more visible in the lower panel
of the figure where the residuals to the best fit function are shown.
The wiggles are quantified by a fit in the energy range
between 100\,GeV and 7\,TeV, well before a possible pile-up in the
spectrum arises. The choice of the fit range is made in order not to bias the
result by the level of the EBL above 10\,$\mu$m, to which  the VHE spectra are
very sensitive due to a super exponential dependency of the attenuation with
the wavelength in that range.
Using the correct EBL level to reconstruct the intrinsic spectrum,
the intrinsic spectrum is well described by a smooth function (Fig.~\ref{Fig:MidEnergyExampleOfMethod}, middle lower panel).
Instead, when using a "wrong" scaled EBL density characteristic deviations
(wiggles) from a smooth function are visible (Fig.~\ref{Fig:MidEnergyExampleOfMethod}, lower panel). The wiggles are quantified by a reduced $\chi^{2}$ value of 51.4/12, corresponding
to a fit probability of $7.9 \cdot 10^{-7}$.
The small fit probability for the assumed scaled EBL level
implies (under used assumptions) significant presence of the unphysical wiggles and,
therefore, an exclusion of that particular EBL realization.
For comparison, HEGRA measured spectrum is also de-absorbed using the same two EBL models and the results are shown by grey filled and open circles in the upper panel. The residuals to the best fits are shown in the middle upper panel. The difference between the two curves (represented by grey filled and open circles) is not significant and is difficult to see. The similar reduced $\chi^{2}$ values of 13.6/8 and 14.0/8 underline that the HEGRA measurement was not sensitive to the method.

\begin{figure}[tbp]
\centering
\includegraphics[width=0.65\textwidth]{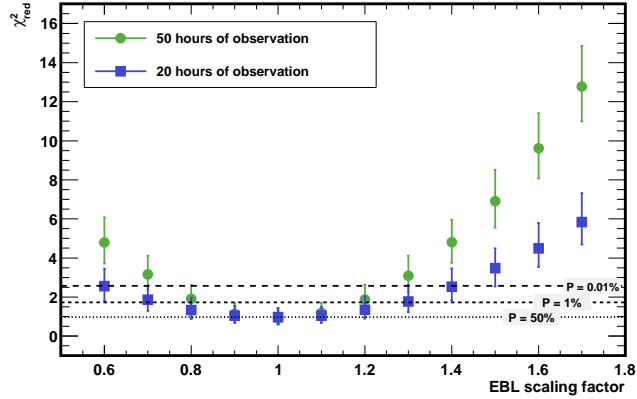}
\caption{
Quantitative results from the search for EBL signatures in the mid-VHE using
the energy spectrum of Mkn\,501.  Shown are the reduced $\chi^{2}$ values from
the fits to the reconstructed spectra of Mkn\,501 as a function of EBL scaling
factor. Blue filled squares and green filled circles show the expected result
for 20 and 50 hours of NGTCS observations, respectively. The black horizontal lines
correspond to fit probabilities as labeled.}
\label{Fig:MidEnergyStatistics}
\end{figure}
 
\paragraph{Results of the analysis}
This procedure is repeated 1000 times for every scaled EBL density in order
to achieve a solid statistical mean and a $1\,\sigma$ (68\%) coverage of the reduced $\chi^{2}$ values. The results are shown in Fig.~\ref{Fig:MidEnergyStatistics} for two different NGCTS exposures (20 and 50 hours) plotting the mean reduced $\chi^{2}$ values including 68\% error bars versus the EBL scaling factor. As a result one can see a parabola-like curve with the minimum at EBL scaling factor of  1.0, which is via construction the correct EBL model.
The expected mean reduced $\chi^{2}$ values are shown by the blue filled squares and the green filled circles for the exposures of 20\,hr and 50\,hr, respectively. The corresponding fit probabilities of P=50\%, 1\% and 0.01\% are shown by the dotted, short-dashed and long-dashed lines, respectively.
An EBL scaling is considered to be excluded when the mean reduced $\chi^{2}$ value including its 68\% error exceeds the $\chi^{2}$ value for P=0.01\%.
For the case constructed here this means that in case of 50 hours observation,
the EBL scalings below 0.65 and above 1.35 are excluded. In the case of 20 hours observation,
the EBL scalings above 1.50 are excluded.

\paragraph{Caveats of the method}
The method presented above is not only sensitive in constraining the EBL
density but it is also a first attempt to resolve the actual EBL level. 
Still, the method has several caveats:
(i) There is a possibility that the intrinsic VHE spectrum is not smooth for some AGNs. 
        For example, several emission regions or several
        generations of charged emitting particles (e.g., electrons) may produce
        spectral features which would mimic the signature from a wrong EBL model. 
        Though neither the AGNs observed at low energies with Fermi/LAT,
       nor the extragalactic VHE $\gamma$-ray sources observed so far with IACTs have
       shown structured spectra, such wiggles can be interpreted as the result of a wrong EBL model only if found to exist in several spectra of different sources consistently.
(ii) The spectral signature shown in
        Fig.~\ref{Fig:MidEnergyExampleOfMethod} can only be
        recognized well after a long exposure of a source being in a flare state. The
        data  of the Mkn\,501 flare from 1997 used for the study remains unique so far.  This means that
        it cannot be expected that a study with such precision can be done on many
        sources with future NGCTS. 
        Still, the indirect EBL measurement based on few strong flares will be of high importance.
(iii) The limited energy resolution of IACTs ($\sim$10\%) could affect the possibility of detecting such a signature. The overall shape of the attenuation effect is broad enough in energy to not strongly be affected by the smoothing of an energy resolution of $<10$\%. For nearby sources, however, the wiggle structure produced by the wrongly assumed EBL model is only of order of a few per cent of the total flux level. An additional spread introduced by the spill-over effects due to events with mis-reconstructed energy may enhance or alter the wiggles. Therefore, an efficient and precise spectral unfolding technique will be key to overcome these problems for nearby sources. For more distant sources the attenuation signatures get quickly stronger. It will be, therefore, highly interesting to study a strong flaring state from a source at redshift $\gtrsim0.1$ with the NGCTS.

\section{Summary \& Conclusions}\label{conclusions}

In this paper the potential of a Next Generation Cherenkov Telescope System (NGCTS)  to study the EBL through observations of VHE spectra from distant sources is explored. In the focus of the study lies the energy range between 40\,GeV and 10\,TeV, where a factor 10 improvement in sensitivity over current generation experiments is expected. Two different methods are investigated:  (i) utilizing the unabsorbed part of the VHE spectrum in the energy range 40-100\,GeV and (ii) searching for attenuation modulation signatures at energies between 100\,GeV to 7\,TeV. While some caveats, like e.g. the exact shape of the intrinsic spectra, do exist, overall the two methods show promising results, clearly go beyond what is possible with current generation instruments.

EBL attenuation is a global phenomenon that affects the spectra of all sources in the same way. It is therefore quite natural to expand the studies on the EBL by combining the results from different methods, sources or source flux states. This can e.g. be done by utilizing log-likelihood methods to combine the constraints from different sources and methods. In addition only the low and mid-VHEs have been investigated. At energies $>10$\,TeV further signatures from EBL attenuation ("cut-off") are expected, which have been used extensively in the past to derive limits on the EBL density in the FIR.

In this paper only data from a single instrument - a NGCTS - and minimal assumptions about intrinsic spectrum have been used to derive constraints on the EBL density. AGNs are observed all across the electromagnetic spectrum and the theory predicts connections between the observations at different energies. A wealth of multi-wavelength information is therefore accessible, which - in combination with a theoretical model - can be used to constrain the spectrum at VHE. A first model-dependent approach to determine the EBL level using a consistent SED modeling of detected blazars was discussed recently by \cite{mankuzhiyil:2010a}. While a detailed discussion of this topic is beyond the scope of this paper, as an example it should only be mentioned the power of combining the \Fe observations at MeV to GeV energies with observations from an NGCTS: such observations will cover over 9 decades (!) in energy with high precision.

The sources under study show strong flux variability in the VHE band. The high sensitivity of a NGCTS will enable to study the energy spectrum on very short time-scales in great detail. The time resolution will enable stronger constraints on the theoretical modeling of the source spectra since the experimental data will be available on shorter scales than the relevant changes in the emission regions. This will further help to disentangle source intrinsic effects and EBL attenuation.

In this study only the VHE spectra of a few selected known sources have been investigated. The NGCTS will bring new discoveries. With the high sensitivity in the 20 to 100\,GeV range it will be possible to detect fainter sources and sources at higher redshift. A (sufficiently) large sample of sources will serve a two-fold purpose: (1) help to improve the understanding of the intrinsic source physics and (2) enable statistical studies of the EBL effects, e.g. by investigating the EBL attenuation features versus redshift. This will also enable to not only probe the present day EBL but to study its evolution with redshift. How such a study can be performed with the help of an NGCTS will be the topic of a second paper.

\section*{Acknowledgments}

{\small
\noindent MR and DM would like to thank M. Persic and M. Tluczykont for the careful reading of the manuscript and useful comments.
MR and DM acknowledge fruitful discussion on CTA performance with J. Hinton and D. F. Torres. The authors would like to thank the referee for helpful comments and suggestions which improved the paper.
DM acknowledges the support by a Marie Curie Intra European Fellowship within
the 7th European Community Framework Programme. 
}

\appendix
 
\section{References for VHE spectral data}

\begin{table}[h]
\begin{center}
\begin{tabular}{ccc}
\hline \hline
Source name & Redshift & Reference \\ \hline
Mkn\,501 & 0.034 & \cite{aharonian:1999b} \\
PKS\,2155-304 & 0.116 & \cite{aharonian:2005:hess:pks2155mwl}, \cite{aharonian:2007:hess:pks2155:bigflare} \\
1ES\,1101-232 & 0.188 & \cite{aharonian:2006:hess:ebl:nature} \\
\hline \hline
\end{tabular}
\caption{\label{Table:VHESpectra} VHE spectra used in this publication.}
\end{center}
\end{table}
 


\def\Journal#1#2#3#4{{#4}, {#1}, {#2}, #3}
\def\NAT{Nature}
\def\AAA{A\&A}
\def\ApJ{ApJ}
\def\AJ{Astronom. Journal}
\def\Aph{Astropart. Phys.}
\def\ApJS{ApJSS}
\def\MNRAS{MNRAS}
\def\NIM{Nucl. Instrum. Methods}
\def\NIMA{Nucl. Instrum. Methods A}
\def\aj{AJ}%
\def\actaa{Acta Astron.}%
\def\araa{ARA\&A}%
\def\apj{ApJ}%
\def\apjl{ApJ}%
\def\apjs{ApJS}%
\def\ao{Appl.~Opt.}%
\def\apss{Ap\&SS}%
\def\aap{A\&A}%
\def\aapr{A\&A~Rev.}%
\def\aaps{A\&AS}%
\def\azh{AZh}%
\def\baas{BAAS}%
\def\bac{Bull. astr. Inst. Czechosl.}%
\def\caa{Chinese Astron. Astrophys.}%
\def\cjaa{Chinese J. Astron. Astrophys.}%
\def\icarus{Icarus}%
\def\jcap{J. Cosmology Astropart. Phys.}%
\def\jrasc{JRASC}%
\def\mnras{MNRAS}%
\def\memras{MmRAS}%
\def\na{New A}%
\def\nar{New A Rev.}%
\def\pasa{PASA}%
\def\pra{Phys.~Rev.~A}%
\def\prb{Phys.~Rev.~B}%
\def\prc{Phys.~Rev.~C}%
\def\prd{Phys.~Rev.~D}%
\def\pre{Phys.~Rev.~E}%
\def\prl{Phys.~Rev.~Lett.}%
\def\pasp{PASP}%
\def\pasj{PASJ}%
\def\qjras{QJRAS}%
\def\rmxaa{Rev. Mexicana Astron. Astrofis.}%
\def\skytel{S\&T}%
\def\solphys{Sol.~Phys.}%
\def\sovast{Soviet~Ast.}%
\def\ssr{Space~Sci.~Rev.}%
\def\zap{ZAp}%
\def\nat{Nature}%
\def\iaucirc{IAU~Circ.}%
\def\aplett{Astrophys.~Lett.}%
\def\apspr{Astrophys.~Space~Phys.~Res.}%
\def\bain{Bull.~Astron.~Inst.~Netherlands}%
\def\fcp{Fund.~Cosmic~Phys.}%
\def\gca{Geochim.~Cosmochim.~Acta}%
\def\grl{Geophys.~Res.~Lett.}%
\def\jcp{J.~Chem.~Phys.}%
\def\jgr{J.~Geophys.~Res.}%
\def\jqsrt{J.~Quant.~Spec.~Radiat.~Transf.}%
\def\memsai{Mem.~Soc.~Astron.~Italiana}%
\def\nphysa{Nucl.~Phys.~A}%
\def\physrep{Phys.~Rep.}%
\def\physscr{Phys.~Scr}%
\def\planss{Planet.~Space~Sci.}%
\def\procspie{Proc.~SPIE}%
          


\bibliographystyle{elsarticle-num}





\end{document}